\documentclass[a4paper,11pt]{article}
\usepackage[utf8]{inputenc}
\usepackage{jheppub} 
\usepackage{orcidlink}

\usepackage{subfigure}
\usepackage{mathtools}
\usepackage{amsmath}
\usepackage{amsfonts}
\usepackage{amssymb}
\usepackage{bbold}
\usepackage{bm}
\usepackage{ esint }
\usepackage{orcidlink}
\usepackage{tabularx}

\usepackage{tikz}
\usetikzlibrary{tikzmark,decorations.pathreplacing}

\newcommand{\exclude}[1]{{}}
\long\def\exclude#1{}

\title{A generalized study of linear electromagnetic cascades in astrophysical sources}

\author[a]{Damiano F.\ G.\ Fiorillo \orcidlink{0000-0003-4927-9850}} 
\affiliation[a]{Deutsches Elektronen-Synchrotron DESY,
Platanenallee 6, 15738 Zeuthen, Germany}

\author[a]{Federico Testagrossa
\orcidlink{0009-0000-9401-1971}}

\author[a]{Chengchao Yuan
\orcidlink{0000-0003-0327-6136}}

\author[b,c]{Maria Petropoulou
\orcidlink{0000-0001-6640-0179}}
\affiliation[b]{Department of Physics, National and Kapodistrian University of Athens, University Campus Zografos, GR 15784, Athens, Greece }
\affiliation[c]{Institute of Accelerating Systems \& Applications, University Campus Zografos, Athens, Greece}

\author[a]{Walter Winter
\orcidlink{0000-0001-7062-0289}}

\abstract{High-energy gamma rays can trigger electromagnetic cascades via pair production on ambient photons, reprocessing their energy to lower frequencies. A classic example is the cascade from the gamma rays produced by ultra-high-energy cosmic rays in extragalactic photon fields, whose universal spectral shape was first described by Berezinsky in the 1970s. Recently, internal cascades, developing within the gamma-ray sources themselves, have gained a prominent role, as the IceCube data suggest that most detected neutrinos originate in gamma-ray-opaque environments. We analyze under what conditions these internal cascades can approach a universal spectrum. Since the Berezinsky treatment breaks down if synchrotron losses dominate, we present a generalized theory incorporating synchrotron-dominated cascades. We show the emergence of universal cascade spectrum among various examples of high-energy sources containing non-thermal cosmic rays, and discuss the conditions for its appearance.}

\begin{document}
\maketitle
\flushbottom

\section{Introduction}
\label{sec:intro}

When a very high-energy gamma-ray $\gamma$ passes through an environment containing low-energy photons $\gamma_t$, the latter act as a target for pair production $\gamma +\gamma_t\to e^+ + e^-$. This initiates a so-called electromagnetic cascade, where the original energy of the gamma-ray is reprocessed into a continuum spectrum that extends down to much lower energies. Historically, the role of the cascade was first appreciated in the context of high-energy gamma rays propagating in the extragalactic space~\cite{Berezinsky:1975zz,Strong:1974tp}, permeated by the Cosmic Microwave Background (CMB) and the Extragalactic Background Light (EBL). These target photon fields make the extragalactic space opaque to gamma rays above tens of TeV~\cite{Gould:1966pza,Stecker:2005qs,Kneiske:2010pt,Dominguez:2010bv,Dwek:2012nb,Stecker:2016fsg,Franceschini:2017iwq}, and the resulting cascade extending to lower energies is the main observational signature of the original high-energy emission (see, e.g., Refs.~\cite{Murase:2013rfa,Murase:2015xka,Kistler:2015ywn,Bechtol:2015uqb,Chang:2014hua,Tamborra:2014xia,Chang:2016ljk,Xiao:2016rvd,Ambrosone:2020evo,Capanema:2020rjj,Capanema:2020oet} for a collection of constraints on neutrino sources from the putative cascade from their accompanying gamma rays).

The cascade evolution through the CMB and EBL can now be followed numerically, but a theory of the final photon spectrum was already described in the seminal work of Ref.~\cite{Berezinsky:1975zz} and refined in later studies. A clear presentation is given in Ref.~\cite{Berezinsky:2016feh}, whose key assumption is that high-energy gamma rays propagate long enough to reach a saturated spectrum, with all energy reprocessed to lower energies in steady state -- a {\bf fully developed cascade}. This requires interaction timescales for high-energy gamma rays and electrons to be much shorter than their propagation times. Magnetic fields are also neglected in this approach. The resulting theory agrees well with simulations for sufficiently distant sources.

An alternative is an \textit{internal} cascade, developing within the astrophysical accelerators of cosmic rays (CRs). Interest in such gamma-ray–opaque sources was spurred by IceCube’s discovery~\cite{IceCube_2013} of a diffuse 1–10~TeV neutrino flux without corresponding gamma-ray emission in Fermi-LAT data. This rules out neutrino production in sources transparent to $\gamma\gamma$ interactions, which would yield a bright GeV cascade~\cite{Murase:2013rfa,Murase:2015xka,Capanema:2020oet,Capanema:2020rjj}. In contrast, $\gamma$-ray–opaque sources -- such as AGN cores, where strong coronal X-ray fields attenuate $\gamma$ rays above tens of MeV -- avoid these constraints, a possibility further strengthened by the recent discovery of a TeV neutrino signal from the gamma-ray-opaque Seyfert galaxy NGC~1068~\cite{IceCube-NGC1068}.

For these opaque gamma-ray sources, the electromagnetic signature is entirely determined by the cascade signal. Yet, the standard theory~\cite{Berezinsky:2016feh} developed for extragalactic cascades does not apply to such environments. The case of AGN coronae offers a simple counterexample; if the strong coronal emission is powered by magnetic dissipation, either via stochastic acceleration in magnetized turbulence~\cite{Murase:2019vdl,Fiorillo:2024akm} or via magnetic reconnection~\cite{Fiorillo:2023dts, Karavola:2024uui}, the magnetic field energy density is larger than the radiation energy density, so synchrotron losses, irrelevant for extragalactic cascades, are a primary source of electron energy losses. This is indeed confirmed by numerical simulations of the cascades~\cite{Fiorillo:2023dts, Karavola:2024uui}. 

Beyond AGN coronae, hadronic cascades, triggered from hadronic CRs which inject high-energy gamma rays, may occur in several astrophysical environments. In Gamma-Ray Bursts (GRBs) -- brief flashes of gamma rays typically from stellar collapse or neutron star mergers -- cascades from internal dissipation have been proposed as the origin of hard power-law components in bright bursts~\cite{FermiGBM:2009ubu,Racusin:2008pd,Asano:2010gs}, reprocessing energy into the GeV–TeV band and probing CR acceleration~\cite{Bottcher:1998qi,Asano:2007my,Asano:2009ta}. Since both $\gamma$ rays and neutrinos arise from the same hadronic interactions, the cascade flux directly constrains neutrino production~\cite{Petropoulou:2014awa}.

In blazars -- AGN with relativistic jets toward Earth -- lepto-hadronic models predict subdominant cascades from hadronic $\gamma$ rays interacting with the jet’s leptonic field. For TXS~0506+056, the first source linked to a high-energy IceCube neutrino~\cite{IC_TXS_2017}, the accompanying X-ray cascade severely limits the neutrino yield~\cite{Ansoldi2018,Keivani:2018rnh,Liu2019,Gao2019,Cerruti2019}. Similar constraints also exist for the reported excess of neutrinos in the direction of TXS~0506+056 in 2014/15, which make an astrophysical explanation for these events quite challenging~\cite{Murase2018, Reimer2019, Rodrigues:2018tku, Petropoulou:2019zqp}. A non-jetted explanation for the neutrinos from TXS~0506+056 faces similar and even greater challenges~\cite{Fiorillo:2025cgm}.

Tidal disruption events (TDEs)—where a star is torn apart by a supermassive black hole—have also been tentatively associated with IceCube neutrinos\footnote{These TDEs lie outside the 90\% angular uncertainty of reconstructed IceCube tracks~\cite{Zegarelli:2025vnq}, weakening the correlations, but Fermi upper limits remain informative.}, including AT2019dsg~\cite{Stein:2020xhk}, AT2019fdr~\cite{Reusch:2021ztx}, AT2019aalc~\cite{vanVelzen:2021zsm}, and AT2021lwx~\cite{Yuan:2024foi}. In these systems, hadronic models predict cascaded $\gamma$ rays from proton interactions with accretion-disk and outflow photon fields~\cite{Murase:2020lnu,Winter:2020ptf,Liu:2020isi,Winter:2022fpf,Yuan:2023cmd}, providing complementary probes of neutrino production.

This wide variety of environments, all exhibiting radiation from electromagnetic cascades, begs the question: is the resulting spectrum the same, independently of the details of the environment, just as the Berezinsky prediction for extragalactic cascades was independent of the detailed spectrum of CMB and EBL? We tackle this question here, generalizing the Berezinsky theory to include the strong magnetic fields usually present in astrophysical sources. We develop a theory for fully developed and {\bf \textit{linear} electromagnetic cascades}, i.e. the cascade photons and electrons interact only with the fixed target photons, not among themselves, so they act as test particles; in other words, the cascade luminosity scales linearly with the injected gamma-ray luminosity.  We show that a universal spectral shape is generally obtained when synchrotron radiation is the dominant energy loss channel for pairs, and clarify the conditions under which this universal prediction might not apply.

We start with the most idealized cases, which allow a complete analytic understanding. Thus, in Secs.~\ref{sec:ic_cascade} and~\ref{sec:synchrotron_cascade}, we discuss the case of a monochromatic target photon field, assuming inverse Compton (IC) and synchrotron, respectively, as the dominant energy loss for leptons. 
In Sec.~\ref{sec:general_treatment}, we abstract these regimes in their universal features, independent of the specific examples, and discuss more broadly under what conditions the universal cascade can be expected; readers uninterested in technical details may skip directly to this section. In Sec.~\ref{sec:astro_sources}, we consider various benchmarks for typical conditions in high-energy astrophysical sources, including GRBs, AGN coronae, blazars, and TDEs. Finally, in Secs.~\ref{sec:summary} and~\ref{sec:discussion}, we summarize our results and discuss their relevance in the context of high-energy astrophysics. Throughout this work, we use natural Gaussian units, in which $\hbar=c=1$.

\section{Inverse-Compton-dominated regime}\label{sec:ic_cascade}

We first consider the simplest case of electromagnetic cascade, triggered by IC interactions. 
For reference, the notation adopted in this work, comprehensive of all the kinematic quantities used, is summarized in Table~\ref{tab:notation}, though we will introduce each quantity throughout the text at their first appearance. In this section, we focus on a monochromatic target photon field at a fixed target energy $\varepsilon_t$, and assume a vanishing magnetic field to avoid synchrotron losses. The threshold for pair production is therefore $\varepsilon_{\gamma, \rm thr}\simeq m_e^2/\varepsilon_t$; we will assume fully developed cascades, such that all photons above this threshold can efficiently produce pairs. The cascade is then triggered by the injection of high-energy gamma rays at a fixed energy $\varepsilon_{\gamma, \rm he}\gg \varepsilon_{\gamma, \rm thr}$. The target photon field has an energy density $u_t$, while the high-energy gamma rays are injected with a luminosity $L_{\gamma, \rm he}$. To discuss the cascade development, we separately consider the optically thick ($\varepsilon_\gamma> \varepsilon_{\gamma, \rm thr}$) and thin ($\varepsilon_\gamma< \varepsilon_{\gamma, \rm thr}$) energy range. In these theoretical calculations, we will neglect the effect of particle escape, focusing on the optimal regime for cascade formation of extreme optical thickness. In the numerical examples, we will show explicitly the effect of a low-energy escape on the cascade.

\begin{table}[h!]
\centering
\renewcommand{\arraystretch}{1.3}
\begin{tabular}{|c|l|c|}
\hline
\textbf{Symbol} & \textbf{Description} & \textbf{Units} \\
\hline
\multicolumn{3}{|c|}{\textbf{Kinematic Quantities}} \\
\hline
\( \varepsilon_e \), \( \varepsilon_p \), \( \varepsilon_\gamma \), \( \varepsilon_t \) & Lepton, proton, photon, and target photon energy & eV \\
\( \gamma_e = \varepsilon_e / m_e \), \( \gamma_p = \varepsilon_p / m_p \) & Lepton and proton Lorentz factor & – \\
\hline
\multicolumn{3}{|c|}{\textbf{Distribution Functions}} \\
\hline
\( n_e(\varepsilon_e) = \frac{dN_e}{d\varepsilon_e} \) & Lepton energy distribution & eV$^{-1}$  \\
\( n_p(\varepsilon_p) = \frac{dN_p}{d\varepsilon_p} \) & Proton energy distribution & eV$^{-1}$  \\
\( n_\gamma(\varepsilon_\gamma) = \frac{dN_\gamma}{d\varepsilon_\gamma} \) & Photon energy distribution & eV$^{-1}$  \\
\( n_t(\varepsilon_t) =\frac{dN_t}{d\varepsilon_t} \) & Target photon energy distribution & eV$^{-1}$ \\
\hline
\multicolumn{3}{|c|}{\textbf{Source and Field Properties}} \\
\hline
\( R \) & Radius of the emission region & cm \\
\( B \) & Magnetic field strength & G \\
\( t_{\rm esc} = R /c \) & Escape timescale & s \\
\hline
\multicolumn{3}{|c|}{\textbf{Spectral properties}} \\
\hline

\( \varepsilon_{\gamma, \rm thr} \) & Photon energy threshold for pair production & eV \\
\( \varepsilon_{\gamma, \rm he} \) & Photon energy of gamma-ray injection & eV \\
\( \varepsilon_{e, \rm thr}=\varepsilon_{\gamma, \rm thr}/2 \) & Energy below which pair injection stops & eV \\
\( \varepsilon_{e, \rm esc} \) & Energy below which leptons escape & eV \\
\( \hbar \omega_B = \hbar e B/m_e c \) & Cyclotron energy & eV \\
\( s_e,\, s_p,\, s_\gamma, s_t\) & Spectral index ($n(\varepsilon)\propto \varepsilon^{-s}$) & - \\
\hline
\multicolumn{3}{|c|}{\textbf{Luminosities and Energy Densities}} \\
\hline
\( \varepsilon L_\varepsilon = \varepsilon^2 n(\varepsilon) / t_{\rm esc} \) & Spectral luminosity of particles & erg s$^{-1}$ \\
\( u_t \) & Target photon energy density & erg cm$^{-3}$ \\
\( L_{\gamma, \rm he} \) & Integrated high-energy gamma-ray luminosity & erg s$^{-1}$ \\
\( u_B = B^2 / (8\pi) \) & Magnetic field energy density & erg cm$^{-3}$ \\
\( b(\varepsilon)=-d\varepsilon/dt\) & Energy loss rate & eV s$^{-1}$ \\
\hline
\end{tabular}
\caption{Summary of notation used for particle distributions, kinematics, and physical parameters of the emission region. In this table only, we temporarily restore standard Gaussian (i.e. non-natural) units for clarity.}
\label{tab:notation}
\end{table}

\subsection{Theoretical calculation}

For $\varepsilon_\gamma>\varepsilon_{\gamma,\rm thr}$, a steady state is reached by the balance between pair production $\gamma+\gamma_t\to e^++e^-$ and IC scattering $e^\pm + \gamma_t \to e^\pm + \gamma$; the target photons are denoted by a suffix $t$, and act as a fixed background -- due to the linearity of cascade it is unaffected by the latter. In each such reaction, a primary particle ``splits'' into two particles with comparable energies, so that the initial gamma-ray energy injected at high energies is cascaded down by a continuous splitting into more particles. We call this the \textbf{equal-reproduction} regime, since its characteristic feature is the splitting of each particle into two particles with comparable energies. To determine the resulting steady spectrum of gamma rays and pairs, we adopt the delta-function approximation, in which in $\gamma+\gamma_t\to e^++e^-$ each gamma-ray produces exactly two leptons with equal energies, and similarly in $e^\pm+\gamma_t\to e^\pm+\gamma$ the two final particles have the same energy. In this approximation, the balance equations for gamma rays and leptons are

\begin{equation}\label{eq:balance_inverse_compton}
\begin{aligned}
\frac{\partial n_e(\varepsilon_e)}{\partial t} 
&= 
\underbrace{4n_\gamma(2\varepsilon_e)\Gamma_{\gamma\gamma}(2\varepsilon_e)}_{\text{\scriptsize$\gamma+\gamma_t \to e^+ + e^-$}} 
+ \underbrace{2n_e(2\varepsilon_e)\Gamma_{\rm IC}(2\varepsilon_e)}_{\text{\scriptsize$e+\gamma_t \to e+\gamma$}} 
- \underbrace{n_e(\varepsilon_e) \Gamma_{\rm IC}(\varepsilon_e)}_{\text{\scriptsize$e+\gamma_t \to e+\gamma$}} = 0, \\
\frac{\partial n_\gamma(\varepsilon_\gamma)}{\partial t} 
&= 
\underbrace{2n_e(2\varepsilon_\gamma)\Gamma_{\rm IC}(2\varepsilon_\gamma)}_{\text{\scriptsize$e+\gamma_t \to e+\gamma$}} 
- \underbrace{n_\gamma(\varepsilon_\gamma)\Gamma_{\gamma\gamma}(\varepsilon_\gamma)}_{\text{\scriptsize$\gamma+\gamma_t \to e^+ + e^-$}} 
+ Q_\gamma = 0.
\end{aligned}
\end{equation}
We have highlighted the reaction to which each term is associated; the factors of 2 
can be understood by noting that in the reaction $e+\gamma_t\to e+\gamma$, where the final-state photon has an energy $\varepsilon_\gamma=\varepsilon_e/2$,
the number of pairs in the initial state is $n_e(2\varepsilon_e) 2d\varepsilon_e$. In the first term for $\gamma+\gamma_t\to e^++e^-$, the factor 4 also accounts for the two leptons produced in the reaction, e.g., $\varepsilon_{\gamma}n_\gamma d{\varepsilon_{\gamma}}=\varepsilon_{e}n_e d{\varepsilon_{e}}|_{\varepsilon_\gamma = 2\varepsilon_e}$.
We also introduce a source term $Q_\gamma$, which in our case is monochromatic at $\varepsilon_{\rm high}$ and so vanishes for most of the energy range of the cascade.
Here $\Gamma_{\gamma\gamma}(\varepsilon_\gamma)$ is the interaction rate for a photon with energy $\varepsilon_\gamma$ to undergo pair production, while $\Gamma_{\rm IC}(\varepsilon_e)$ is the Inverse Compton interaction rate for a lepton with energy $\varepsilon_e$. Both are determined in the Klein-Nishina regime, i.e. when the center-of-mass energy is much larger than the electron rest mass, since $\varepsilon_{e}$ and $\varepsilon_\gamma$ are both larger than $\varepsilon_{\gamma, \rm thr}$; in this range, we have, in order of magnitude (e.g. Ref.~\cite{1990MNRAS.245..453C})
\begin{equation}
    \Gamma_{\rm IC}(\varepsilon_e)\sim  \frac{u_t}{\varepsilon_t} \sigma_T \frac{m_e^2}{\varepsilon_t \varepsilon_{e}},
\end{equation}
where $\sigma_T$ is the Thomson cross section, and the same expression for $\Gamma_{\gamma\gamma}$. Here $u_t/\varepsilon_t$ is the number density of target photons, and $\sigma_T m_e^2/\varepsilon_t \varepsilon$ is the Klein-Nishina suppressed cross section.

The equal-reproduction cascade is a well-known regime across many branches of theoretical physics; the first historical appearance is the shower produced by the passage of energetic particles through matter~\cite{Bhabha:1936ytj, Carlson:1937zz, Landau:1937pqt}. Its defining feature is that, across an energy interval $d\varepsilon$, the number of particles (irrespective of whether they are photons or leptons) passing per unit time is twice as large as those passing through the interval $d(2\varepsilon)$, since a particle with energy $2\varepsilon$ splits into two particles with energy $\varepsilon$. Since the rate with which particles pass through the energy interval $d\varepsilon$ is $\Gamma(\varepsilon)$, where $\Gamma(\varepsilon)$ is the IC rate $\Gamma_{\rm IC}(\varepsilon_e)$ for leptons and the $\gamma\gamma$ rate $\Gamma_{\gamma\gamma}(\varepsilon_\gamma)$ for photons, we must have
\begin{equation}
    \Gamma(\varepsilon)n(\varepsilon)d\varepsilon =2 \left[\Gamma(2\varepsilon) n(2\varepsilon) d(2\varepsilon)\right].
\end{equation}
This equation is satisfied by the solution $n(\varepsilon)\Gamma(\varepsilon)\propto \varepsilon^{-2}$. Hence, at steady state, we expect the lepton and photon distribution to possess such a form. Indeed, an explicit substitution shows that Eqs.~\ref{eq:balance_inverse_compton} at steady state are solved by the ansatz
\begin{equation}\label{eq:equal_reproduction_regime}
    n_e(\varepsilon) \Gamma_{\rm IC}(\varepsilon)=2n_\gamma(\varepsilon)\Gamma_{\gamma\gamma}(\varepsilon)\propto \varepsilon^{-2}.
\end{equation}
We are \textit{not} specifying whether the energy is $\varepsilon_e$ or $\varepsilon_\gamma$, since this equation relates the lepton and photon distribution at the same energy. This is the cascade spectrum in the equal-reproduction regime, which in this case is obtained above the Klein-Nishina threshold $\varepsilon>\varepsilon_{\gamma,\rm thr}$. 

Below $\varepsilon_{\gamma, \rm thr}$, photons are unable to produce further pairs, and simply escape from the interaction region. Therefore, below $\varepsilon_{e,\rm thr}=\varepsilon_{\gamma, \rm thr}/2$ no pairs are produced. In this energy range, the pairs produced at higher energies can only cool, in this case due to IC scattering. We will call this the \textbf{cooling-only} regime.
The energy losses are given by
\begin{equation}\label{eq:ic_energy_loss_rate}
    b_{\rm IC}(\varepsilon_e)=\left(-\frac{d\varepsilon_e}{dt}\right)_{\rm IC}=\frac{4}{3}\sigma_T\left(\frac{\varepsilon_e}{m_e}\right)^2 u_t.
\end{equation}
In equilibrium, leptons must satisfy the steady Fokker-Planck equation in energy 
\begin{equation}
    \frac{\partial}{\partial \varepsilon_e}\left[b_{\rm IC}(\varepsilon_e)n_e(\varepsilon_e)\right]=0,
\end{equation}
and therefore $n_e(\varepsilon_e)\propto \varepsilon_e^{-2}$; the pairs have a constant spectral index $s=2$ throughout the energy range. 
Combining this with the pair spectrum above $\varepsilon_{\gamma, \rm thr}$, we find the general expression for the pair spectrum
\begin{equation}\label{eq:electron_spectrum}
    n_e(\varepsilon_e)\propto  \varepsilon_e^{-2}\mathrm{min}\left[\Gamma_{\rm IC}(\varepsilon_e)^{-1},\Gamma_{\rm IC}(\varepsilon_{\gamma, \rm thr})^{-1}\right].
\end{equation}

Regarding the photons, below $\varepsilon_{\gamma,\rm thr}$ they are optically thin and therefore their spectrum is determined by the radiation spectrum from the pairs. The pairs in the equal-reproduction regime, above $\varepsilon_{e,\rm thr}$, produce photons above $\varepsilon_{\gamma, \rm thr}$. Thus, the photons below $\varepsilon_{\gamma, \rm thr}$ are entirely produced by pairs in the cooling-only regime. The photon spectrum is determined by a balance between the escape rate and the injection from IC scattering; using the delta-function approximation for the IC injection, such that each lepton with energy $\varepsilon_e$ produces a photon with energy $\varepsilon_\gamma=4\varepsilon_t (\varepsilon_e/m_e)^2/3$, we find
\begin{equation}\label{eq:balance_transparency_regime}
    \frac{\partial n_\gamma(\varepsilon_\gamma)}{\partial t}=-\frac{n_\gamma(\varepsilon_\gamma)}{t_{\rm esc}}+\frac{u_t \sigma_T m_e}{2\varepsilon_t}\sqrt{\frac{3}{4\varepsilon_t \varepsilon_\gamma}}n_e\left(m_e\sqrt{\frac{3\varepsilon_\gamma}{4\varepsilon_t}}\right)=0;
\end{equation}
the factor 2 in the IC injection term comes from the derivative $\partial \varepsilon_\gamma/\partial \varepsilon_e$ after integrating the delta function.
This equation gives the gamma-ray spectrum below $\varepsilon_{\gamma, \rm thr}$ (transparency range), while Eq.~\ref{eq:equal_reproduction_regime} gives the gamma-ray spectrum above $\varepsilon_{\gamma, \rm thr}$. Notice that in this energy range the spectrum is much smaller than in the transparency range, by a factor $\Gamma_{\gamma\gamma} t_{\rm esc}\gg 1$, namely the source opacity to pair production; this is due to attenuation, that reduces dramatically the amount of gamma rays in the opaque regime. Therefore, what we are truly interested in is the cascade spectrum in the transparency regime, for $\varepsilon_\gamma<\varepsilon_{\gamma, \rm thr}$; using Eqs.~\ref{eq:electron_spectrum} and~\ref{eq:balance_transparency_regime}, we find $n_\gamma(\varepsilon_\gamma)\propto \varepsilon_\gamma^{-3/2}$. The value of the normalization constant can be found by imposing energy conservation; in the limit $\Gamma_{\gamma\gamma} t_{\rm esc}\gg 1$, essentially all of the energy is emitted in the form of gamma rays in the transparency range, which must therefore carry the entire energy originally injected in high-energy gamma rays. Therefore, we must have
\begin{equation}\label{eq:analytic_ic_cascade}
    n_\gamma(\varepsilon_\gamma)=\frac{L_{\gamma, \rm he} t_{\rm esc}}{2\sqrt{\varepsilon_{\gamma,\rm thr}}}\varepsilon_\gamma^{-3/2}.
\end{equation}
The prediction of $s_\gamma=3/2$ is the same as the original theory of Ref.~\cite{Berezinsky:2016feh}.

\subsection{Numerical testing}

\begin{figure*}
    \includegraphics[width=\textwidth]{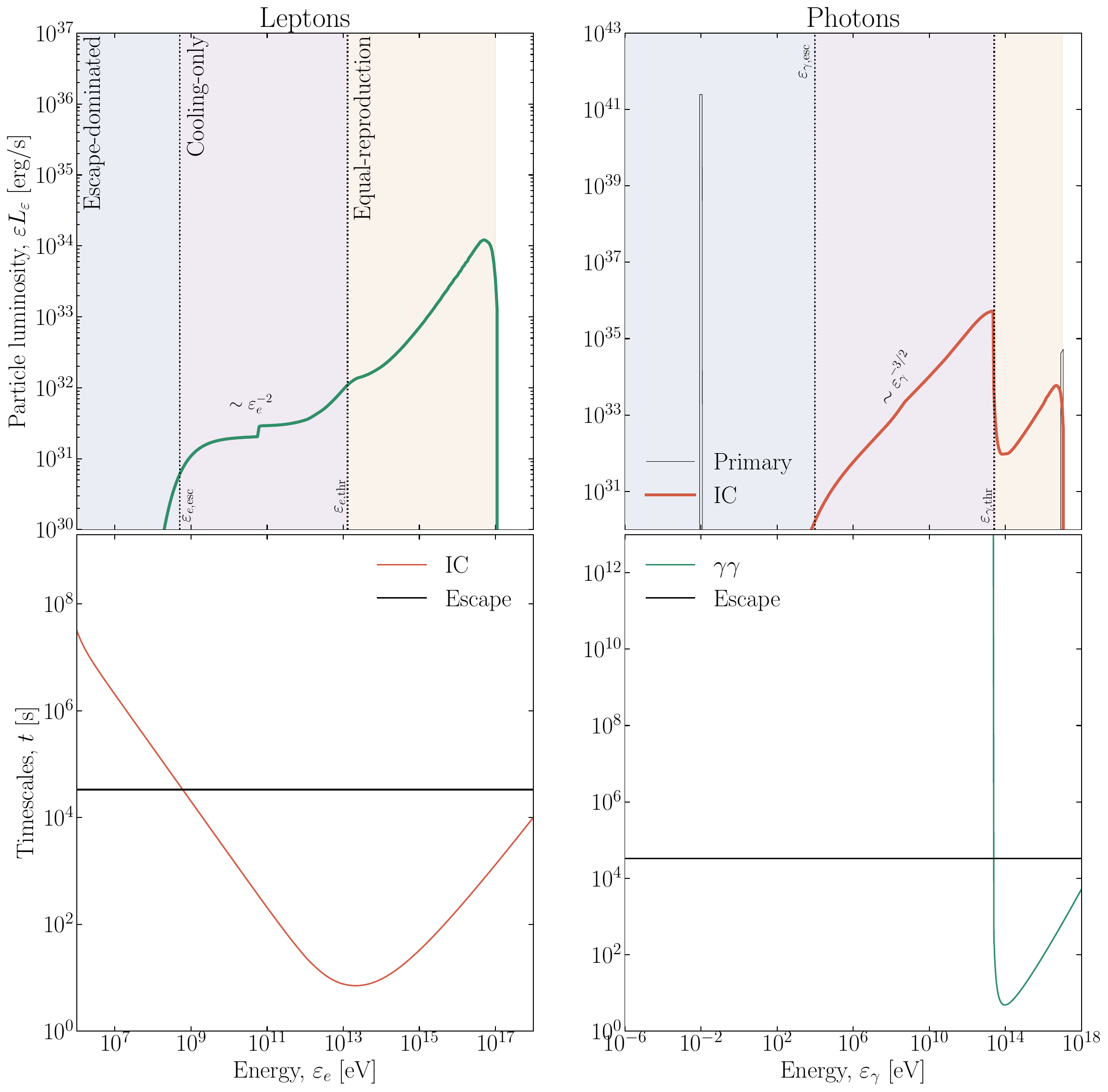}
    \caption{\textbf{IC-dominated cascade for a monochromatic photon target.} Lepton (left) and photon (right) distributions within the interaction region are shown; the setup is described in the main text. Top: luminosity; bottom: interaction and escape timescales (IC scattering, escape, and $\gamma\gamma$ pair production). Thin black lines denote the primary injected species; thick lines show the cascade components. The different regimes of the pair cascade in the left plot are marked by distinct background colors. In the right plot, the same colors indicate the energy ranges where photon injection is dominated by the corresponding regime.
    In the \textit{equal-reproduction} regime, pairs and photons split in energy equally. In the \textit{cooling-only} regime, no pairs are injected; existing pairs cool with an index $s_e=2$, producing photons with $s_\gamma=3/2$. In the \textit{escape-dominated} regime, pairs escape before cooling significantly, and their spectrum is suppressed.}\label{fig:ic_cascade}
\end{figure*}

To test the above setup, we explicitly simulate a closed system with a population of monochromatic target photons with energy $\varepsilon_{t}=10^{-2}$~eV.  The choice of the reference scale is of no particular significance, and it simply determines the threshold for high-energy gamma-ray absorption $\varepsilon_{\gamma, \rm thr}\simeq m_e^2/\varepsilon_t$. These photons are injected with a luminosity $L_t=10^{41}\,\mathrm{erg/s}$ within a spherical region of radius $R=10^{15}\,\mathrm{cm}$. The corresponding energy density is $u_t=3L_t/4\pi R^2$. We also inject a constant high-energy gamma-ray monochromatic flux at an energy scale $\varepsilon_{\gamma,\rm he}=10^{17}\,\mathrm{eV}$, with a luminosity $L_{\gamma, \rm he}=10^{36}\,\mathrm{erg/s}$. All of the simulations in this work are performed using the \texttt{AM$^3$} software~\cite{Klinger:2023zzv}. Here and throughout this work, we assume a free-streaming escape timescale common to all species\footnote{All our simulations are evolved up to $t_{\rm sim}=7t_{\rm esc}$ to ensure convergence to the steady state.} -- protons, leptons, and photons -- and equal to $t_{\rm esc}=R/c$. Such an energy-independent timescale might also be representative of advective escape or adiabatic expansion, while it would not be representative of diffusive escape which usually happens at an energy-dependent rate.

Fig.~\ref{fig:ic_cascade} shows the resulting cascade spectrum, both for leptons and radiation, as well as the timescales for energy loss and escape of the species. We explicitly highlight the different regimes of the pair cascade using different colors. For photons, the same colors indicate the energy ranges where injection is primarily driven by each respective regime.
At very high energies, above $\varepsilon_{\gamma,\rm thr}$, there is the equal-reproduction regime (in orange). The differential luminosity $\varepsilon L_\varepsilon$ follows precisely the shape of $\Gamma_{\rm IC}^{-1}$ for leptons and $\Gamma_{\gamma \gamma}^{-1}$ for photons, as predicted by Eq.~\ref{eq:equal_reproduction_regime}. Below $\varepsilon_{\gamma,\rm thr}$, pairs enter the cooling-only regime (in purple) with their characteristic $n_e(\varepsilon_e)\propto \varepsilon_e^{-2}$ spectrum\footnote{The slight step observable in the pair spectrum is caused by the matching of different integration methods from the \texttt{AM$^3$}  software in different regimes of optical thickness. A similar step discontinuity is found also in the original \texttt{AM$^3$} paper (see Fig.~2 of Ref.~\cite{Klinger:2023zzv}). Its effect is generally mildened on the photon spectrum, and does not affect our conclusions.}, whereas photons follow the analytical prediction $n_\gamma(\varepsilon_\gamma)\propto \varepsilon_\gamma^{-3/2}$. At low energies, below a critical energy $\varepsilon_{e,\rm esc}$ defined by the equality between escape and loss timescale, the IC process becomes slower than the escape of pairs from the interaction region; we identify this as an \textbf{escape-dominated} regime (in blue), causing a drop in the pair density. Since the typical energy at which a lepton with Lorentz factor $\gamma_e$ radiates photons is $\varepsilon_\gamma\simeq \varepsilon_t \gamma_{e}^2$, we have a similar lower cutoff in the photon spectrum at $\varepsilon_{\gamma, \rm esc}\simeq \varepsilon_t \gamma_{e,\rm esc}^2$, where $\gamma_{e, \rm esc}=\varepsilon_{e,\rm esc}/m_e$.

\section{Synchrotron-dominated regime}\label{sec:synchrotron_cascade}

As discussed in the introduction, while the cascades produced in external environments, e.g. in the propagation through the CMB, are likely not strongly affected by synchrotron losses, this is not the case for internal cascades within astrophysical sources. In this case, we should take the opposite regime in which losses are entirely dominated by synchrotron radiation, which to our knowledge has never been discussed before using a theoretical treatment (see Ref.~\cite{Petropoulou:2013gga} for a discussion of synchrotron-dominated \textit{non-linear} cascades, which lead to a very different phenomenology). Synchrotron emission introduces immediately a new energy scale, the cyclotron frequency for an electron at rest $\omega_B=eB/m_e$.

\subsection{Theoretical calculation}

As before, we distinguish regimes below and above the threshold energy $\varepsilon_{\gamma,\rm thr}$. Photons are injected at a characteristic energy $\varepsilon_{\gamma,\rm he}$, which, unlike the previous case, now plays a role in shaping the cascade. These photons produce pairs of comparable energy, which radiate synchrotron photons up to a maximum energy $\sim \omega_B (\varepsilon_{\gamma,\rm he}/m_e)^2$. The cascade behavior thus depends on whether $\omega_B (\varepsilon_{\gamma,\rm he}/m_e)^2 \gg \varepsilon_{\gamma,\rm he}$, i.e., $\varepsilon_{\gamma,\rm he} \gg m_e^2/\omega_B$, or not. In the former case, synchrotron radiation is in the quantum regime, with each lepton emitting photons of comparable energy, and the classical synchrotron description fails. We focus on the opposite regime, $\varepsilon_{\gamma,\rm he} \lesssim m_e^2/\omega_B$, where classical synchrotron remains valid.

For $\varepsilon_e<\omega_B (\varepsilon_{\gamma, \rm he}/m_e)^2$, photons are injected by synchrotron radiation, while pairs are continuously replenished by pair production. This is a new cascade regime, in which the pairs continuously inject photons, but each photon has an energy much lower than the producing lepton; for a lepton with energy $\varepsilon_e$, the typical photon radiated by synchrotron will have an energy $\varepsilon_\gamma\sim \omega_B (\varepsilon_e/m_e)^2$. Thus, we have a \textbf{soft-radiation} regime; it shares similarities with the equal-reproduction regime -- pairs are replenished continuously, and they produce new photons -- but it differs dramatically because the typical photon energy is much lower than the radiating lepton, i.e., ``soft''.

The cascade is thus determined by a balance between pair production and synchrotron radiation, rather than IC scattering. The energy losses due to synchrotron radiation are parameterized by
\begin{equation}\label{eq:syn_loss}
    b_{\rm syn}(\varepsilon_e)=\left(-\frac{d\varepsilon_e}{dt}\right)_{\rm syn}=\frac{4}{3}\sigma_T\left(\frac{\varepsilon_e}{m_e}\right)^2 u_B,
\end{equation}
where $u_B$ is the magnetic field energy density. In the delta-function approximation, we can assume that all of this energy is radiated at a characteristic frequency $\varepsilon_\gamma=4\omega_B (\varepsilon_e/m_e)^2/3$, where $\omega_B$ is the cyclotron frequency. Numerically, we have
\begin{equation}
    u_B=\frac{B^2}{8\pi}=0.04\;B_G^2\; \mathrm{erg/cm}^3,\; \omega_B=\frac{eB}{m_e}=1.2\times 10^{-8}\;B_G\;\mathrm{eV},
\end{equation}
with $B_G=B/1\;\mathrm{G}$.

The properties of this soft-radiation regime are somewhat more complex than the other ones, due to the non-locality in energy of the dynamics. We discuss in Appendix~\ref{app:soft_radiation_cascade} the corresponding steady-state that is achieved. Here we limit ourselves to summarizing the main result of this calculation, namely that the pair spectrum in the soft-radiation regime is approximately a power law $n_e(\varepsilon_e)\propto \varepsilon_e^{-3}$. The soft-radiation regime stops at low energies of the order of $\varepsilon_{e, \rm thr}$, since pair production becomes impossible there. Therefore, at lower energies the pairs enter the cooling-only regime that we have already discussed in the previous section, with $n_e(\varepsilon_e)\propto \varepsilon_e^{-2}$.

As for the radiation, the photons produced by synchrotron radiation from pairs in the soft-radiation regime, with their spectral index $s_e=3$, must have a spectral index $s_\gamma=(s_e+1)/2=2$. The minimum energy at which these photons from the soft-radiation cascade can extend is $\varepsilon_{\gamma, \rm syn}\simeq \omega_B \gamma_{\rm thr}^2$, where $\gamma_{\rm thr}=\varepsilon_{e,\rm thr}/m_e$. At lower energies, photons are produced by pairs in the cooling-only regime, and therefore have a spectral index $s_\gamma=3/2$. Thus, our overall result for the cascade, after normalizing the spectrum so that it has the correct
\begin{equation}\label{eq:synchrotron_cascade_em}
    n_\gamma(\varepsilon_\gamma)=\frac{L_{\gamma, \rm he}t_{\rm esc}}{2+\log(\varepsilon_{\gamma,\rm thr}/\varepsilon_{\gamma,\rm syn})}\mathrm{min}\left[\left(\frac{\varepsilon_\gamma}{\varepsilon_{\gamma, \rm syn}}\right)^{-3/2},\left(\frac{\varepsilon_\gamma}{\varepsilon_{\gamma, \rm syn}}\right)^{-2}\right].
\end{equation}

We remark again that our predictions for the pair spectrum are only valid for $\varepsilon_e\ll \omega_B (\varepsilon_{\gamma, \rm he}/m_e)^2$, since they assume that synchrotron radiation is produced from pairs within the same energy range of the cascade in a self-consistent way. In the range $\varepsilon_e\gg \omega_B (\varepsilon_{\rm max}/m_e)^2$, there is actually a high-energy cooling-only regime, because pairs radiate synchrotron photons at such low energies that there is no further pair injection. However, in realistic situations (see Sec.~\ref{sec:astro_sources}) we do not find instances of this case, which therefore we do not examine in further detail.

\subsection{Numerical testing}

\begin{figure*}
    \includegraphics[width=\textwidth]{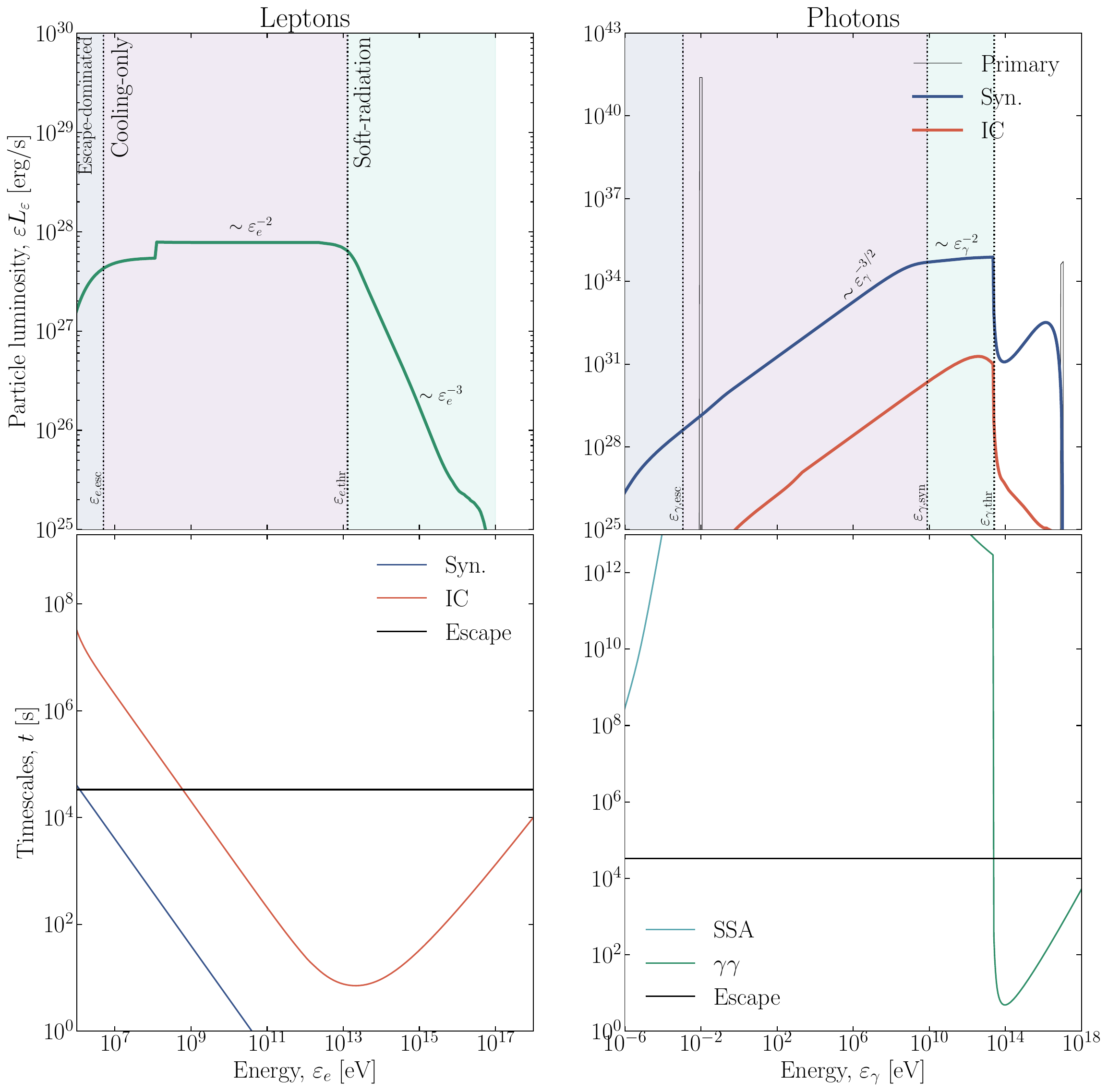}
    \caption{\textbf{Synchrotron-dominated cascade for a monochromatic photon target.} Same as Fig.~\ref{fig:ic_cascade} for a synchrotron-dominated case; the timescales now include also synchrotron radiation and SSA. The setup is described in the main text. At high energies, we have the \textit{soft-radiation} regime, with the photons attenuated by pair production and the pairs settling into the universal $s_e=3$ state, producing photons with $s_\gamma=2$. Below the threshold for pair production, the pairs enter the \textit{cooling-only} regime, with $s_e=2$, and producing photons with $s_\gamma=3/2$. In the \textit{escape-dominated} regime, here achieved for a very narrow low-energy range, pairs escape nearly freely.}\label{fig:syn_cascade}
\end{figure*}

To test this prediction, we now consider the same setup as Fig.~\ref{fig:ic_cascade}, with a monochromatic photon field, but we introduce a magnetic field $B=10^3\,\mathrm{G}$. With such a large field, the energy losses for pairs are dominated by synchrotron losses, and we can therefore validate our predictions. The corresponding electromagnetic cascade is shown in Fig.~\ref{fig:syn_cascade}. 

The cascade is characteristically different from the IC-dominated one. Pairs transition from the soft-radiation regime above $\varepsilon_{e,\rm thr}$ to the pure cooling-only regime below $\varepsilon_{e, \rm thr}$, where pair production is impossible. In turn, the gamma-ray cascade exhibits a broken power-law spectrum, with the break signaling the transition from pairs in soft-radiation regime (at high energies) to pairs in cooling-only regime (at low energies). Finally, also here we identify a low-energy escape-dominated range, albeit in a very narrow interval. In fact, precisely because of the drop in the gamma-ray spectrum due to the escape of pairs, we do not observe any effect due to synchrotron self-absorption (SSA) here, which would set in at lower energies where the photon flux is already suppressed. In later examples where escape is less relevant, as shown in Sec.~\ref{sec:astro_sources}, we find that SSA is usually quite relevant in shaping the photon flux. Overall, the synchrotron-dominated case shows a very good agreement between our analytical prediction and the numerical solution.

\section{When is a universal cascade spectrum expected?}\label{sec:general_treatment}

The two examples we have investigated in Secs.~\ref{sec:ic_cascade} and~\ref{sec:synchrotron_cascade} show the main features of electromagnetic cascades appearing in more realistic cases. In this section, we provide a comprehensive overview of the different regimes of pair cascade formation, whose emergence we have observed in the previous examples, and then discuss under what circumstances these universal cascade regimes, i.e. independent of the specific properties of the target and injection photon spectra, can appear.

\subsection{Summary of cascade regimes}

We have shown the existence of three regimes in the cascade formation:
\begin{itemize}
    \item the \textbf{equal-reproduction} regime, in which pairs and photons split their energy equally among their daughter particles. This is the case in the Klein-Nishina regime, under the reactions $\gamma+\gamma_t\to e^++e^-$ and $e+\gamma_t\to e+\gamma$. In this regime, pairs and photons both obey $n(\varepsilon)\propto \varepsilon^{-2}/\Gamma(\varepsilon)$, where $\Gamma(\varepsilon)$ is the interaction rate of the species. The equal-reproduction regime, which appears in IC-dominated cascades, is interesting in its own right, but usually leads to little consequences because it is difficult to probe observationally; the photons are produced in the optically thick regime and are strongly suppressed;
    \item the \textbf{soft-radiation} regime, in which pairs are injected by $\gamma\gamma$ and cool, but each lepton radiates photons at a frequency much smaller than its energy. This is the reason for the choice of naming it soft-radiation-cascade, since photons are soft -- with much lower energies than the radiating leptons, in the particle-physics sense. This is the case for the bulk of the synchrotron-dominated cascade, in which leptons with a Lorentz factor $\gamma$ radiate at an energy $\omega_B \gamma^2\ll \gamma m_e$. The pairs settle into a power law with spectral index $s_e=3$; in turn, their synchrotron radiation at lower energies, in the optically thin range, has an index $s_\gamma=2$;
    \item the \textbf{cooling-only} regime, in which pairs are not injected and only cool by IC scattering (in the Thomson regime) or synchrotron radiation. In this case, the pairs have an index $s_e=2$, producing low-energy photons,  by either synchrotron radiation or IC scattering, with an index $s_\gamma=3/2$. 
\end{itemize}
In all these regimes, pair cool much faster than they escape, in the same sense as the fast cooling case identified in GRBs~\cite{Sari:1997qe}. However, they are distinguished by whether pairs are injected in the corresponding energy range, and by the typical energy of the photons they radiate.
For a monochromatic target photon spectrum, we have shown that the IC-dominated cascade transitions, from high to low energies, from an equal-reproduction regime to a cooling-only regime. For the synchrotron-dominated cascade, instead, the high-energy range exhibits a soft-radiation cascade, due to the soft nature of the classical synchrotron radiation. In both cases, the cascade is driven by a succession of the elementary regimes we have identified. 

Overall, the ubiquitous appearance of these regimes is responsible for the emergence of a universal spectrum with the cascade photons transitioning from a low-energy state with $s_\gamma=3/2$ to a high-energy state with $s_\gamma=2$; while the details -- break energy, normalization -- may vary with the environment, these spectral indices, as we will see in Sec.~\ref{sec:astro_sources}, appear in most contexts where electromagnetic cascades triggered by high-energy (usually hadronic) injection are present.

\subsection{Limitations of the cascade universality}\label{sec:limitations}

With a more generic understanding of the different cascade regimes, we can now discuss under what conditions they may break down. We will separately discuss different possible factors that can hinder the emergence of a universal cascade.

\subsubsection{Pair escape} 
We have already shown the impact of the escape of low-energy particles from the region of cascade development. As we have seen, this shows up as a drop in the pair spectrum, corresponding to their efficient escape, and causing a corresponding drop in the electromagnetic spectrum. The effect of pair escape appears below a critical pair energy $\varepsilon_{e, \rm esc}$, defined by the condition that the energy-loss timescale and the escape timescale are equal. For the IC-dominated case, this generally depends on the properties of the target photon spectrum. In the synchrotron-dominated case, which is usually the most interesting one for the phenomenology of internal cascades as we will see in Sec.~\ref{sec:astro_sources}, we can obtain an explicit expression for the threshold escape energy by equating the synchrotron energy loss timescale $t_{\rm syn}(\varepsilon_{e, \rm esc})=\varepsilon_{e, \rm esc}/b_{\rm syn}(\varepsilon_{e, \rm esc})$ and the escape timescale $t_{\rm esc}=R$
\begin{equation}
    \varepsilon_{e, \rm esc}=\frac{3 m_e^2}{4\sigma_T R U_B}=1.2\times 10^{13}\, B_G^{-2}\, R_{12}^{-1}\, \mathrm{eV}.
\end{equation}
At lower pair energies, the lepton spectrum drops because pairs manage to escape. In turn, this implies a corresponding drop in the synchrotron photon spectrum at energies below $\varepsilon_{\gamma,\rm esc}\simeq \omega_B (\varepsilon_{e,\rm esc}/m_e)^2$.

\subsubsection{Non-monochromatic target photons}
For the synchrotron-dominated case, the spectral shape of the target photons is inessential, since the energy losses depend only on the magnetic field. However, in the IC-dominated regime, a non-monochromatic target photon field can have a significant impact. The simplest case is that in which the target photons responsible for $\gamma\gamma$ attenuation have an energy $\varepsilon_{t, \rm max}$, but there is a more intense target photon field dominating the IC losses at lower energies $\varepsilon_{t, \rm max}$. A textbook example is the original Berezinsky cascade~\cite{Berezinsky:2016feh}, where the extragalactic background light is responsible for the dominant $\gamma\gamma$ attenuation due to its higher energy range, but the IC losses are dominated by the CMB photons which are much more numerous. In this case, the cascade develops in the soft-radiation regime, similar to the synchrotron-dominated case: photons produce pairs with energy $\varepsilon_e$, which radiate IC photons at a much lower energies $\varepsilon_\gamma\simeq \varepsilon_{t, \rm min}(\varepsilon_e/m_e)^2$. Therefore, the pair cascade has a spectral index $s_e=3$ (soft-radiation) for energies above $\varepsilon_{e, \rm thr}=\varepsilon_{\gamma, \rm thr}$, where $\varepsilon_{\gamma, \rm thr}=m_e^2/\varepsilon_{t, \rm max}$, and a spectral index $s_e=2$ (cooling-only) for lower energies. In turn, the cascade photons from IC scattering off the low-energy target photons with $\varepsilon_{t, \rm min}$ will produce the broken power law spectrum, with $s_\gamma=2$ above the break $\varepsilon_{\gamma, \rm br}\simeq \varepsilon_{t, \rm min}(\varepsilon_{e, \rm thr}/m_e)^2$ and $s_\gamma=3/2$ at lower energies.
In Appendix~\ref{sec:non_monochromatic_ic}, we show explicitly the emergence of this spectrum in a numerical example. 

If the target photon spectrum does not have two characteristically defined energies $\varepsilon_{t, \rm min}$ and $\varepsilon_{t, \rm max}$, then the IC cascade will lose its universality. As we prove in more detail in Appendix~\ref{sec:non_monochromatic_ic}, this can happen when the target photon spectrum is extended over a wide energy range, e.g. with a power-law behavior $n_t(\varepsilon_t)\propto \varepsilon_t^{-s_t}$ 
with $0<s_t<2$. Harder target spectra, with $s_t<0$, can be approximated by a monochromatic spectrum; softer target spectra, with $s_t>2$, can be approximated by a bichromatic spectrum, with the soft component at $\varepsilon_{t, \rm min}$ dominating the IC losses and the hard component at $\varepsilon_{t, \rm max}$ (or more generally the maximal energy at which the target photons make the environment optically thick) dominating the $\gamma\gamma$ reactions.

\subsubsection{Non-monochromatic high-energy injection}

If the primary gamma rays triggering the cascade are injected over an extended energy range, and in particular throughout the range where the cascade develops, the resulting emission will not belong to the universal spectra we have identified. Since the cascade is linear, one might still obtain the resulting electromagnetic spectrum analytically, by superimposing the emission from particles injected at each energy interval. However, such a procedure cannot be performed in a general way, and depends on the specific properties of the injection. Therefore, in these cases the universal cascade prediction fails. 

In real astrophysical sources, this case can appear when the electromagnetic cascade is triggered by leptonic emission. If high-energy leptons are accelerated, they usually inject photons over a wide range of energy in which the cascade develops. Therefore, leptonic-triggered cascades often do not exhibit the typical features we have identified. Instead, hadronic cascades, triggered by gamma rays injected by photohadronic interactions, satisfy our criterion; since the photohadronic efficiency increases with energy, most gamma rays are injected at very high energies, akin to a monochromatic injection, and therefore for a wide energy range we expect the formation of the universal cascade.  One exception is the case of dominant Bethe-Heitler (BH) processes, which inject pairs over a very wide energy range (see, e.g., Refs.~\cite{Kelner:2008ke,Karavola:2024uog}). Thus, in this case, the approximation of gamma rays being injected only at very high energies, where the photohadronic efficiency peaks, breaks down. Therefore, in those cases where BH processes are the dominant injection of non-thermal particles, the universal shape of the cascade might not be recovered.

\section{Electromagnetic cascades in high-energy astrophysical sources}\label{sec:astro_sources}

In this section, we present a collection of examples, motivated by various classes of astrophysical sources, which exhibit the universal cascade prediction we have obtained. This shows more clearly the practical circumstances under which the universal cascade can appear, and its potential effects on the phenomenology of high-energy sources. For some of the cases discussed below, such as GRBs and blazars, the radiation is produced in zones moving at relativistic speeds. Accordingly, the physical quantities and spectra are presented in the comoving frame. For all sources, we use the numerical code \texttt{AM$^3$} to simulate the evolution of the non-thermal particles until they reach a steady state; this includes all the main radiative processes for protons ($p\gamma$ interactions, Bethe-Heitler process, synchrotron radiation, IC scattering), leptons (synchrotron radiation, IC scattering), mesons (synchrotron radiation, IC scattering, decay), photons (pair production, Compton scattering, SSA).

\subsection{AGN coronae}

AGN are powerful non-thermal sources, primarily driven by the accretion onto supermassive black holes. While a fraction of them exhibits a strongly beamed electromagnetic emission within their jets, in this section we focus on non-jetted AGN, which also have the potential for high-energy neutrino and gamma-ray production. While this general idea dates back in time~\citep{Stecker:1991vm}, it has gained new traction in recent years, after the realization that the brightest hotspot of neutrinos in the sky comes from the direction of Seyfert II galaxy, NGC~1068. While this source does exhibit a weak jet, this is unlikely to be connected with neutrino production, since a jet emission would lead to an accompanying gamma-ray emission which is not observed by MAGIC~\citep{MAGIC-UL-NGC1068}. Instead, a large opacity to $\gamma\gamma$ absorption, which is a natural requirement to attenuate the high-energy photons and explain the missing gamma rays, is naturally achieved in the inner regions of the accretion flow, and in particular in the corona. The latter is a compact, hot plasma region located above the accretion disk, believed to be responsible for the hard X-ray emission commonly observed in radio-quiet AGN. Intriguingly, the optical thickness of AGN coronae is also an intrinsic requirement for the dominant sources of the diffuse neutrino flux observed by IceCube in the 1-100~TeV energy range~\citep{Murase:2013rfa,Murase:2015xka,Capanema:2020rjj,Capanema:2020oet}, potentially suggesting a common origin for the whole of this neutrino flux~\cite{Murase:2019vdl,Kheirandish:2021wkm,Padovani:2024tgx,Ambrosone:2024zrf,Fiorillo:2025ehn}.

\begin{table}[h!]
\centering
\begin{tabularx}{\linewidth}{|l|c|c|c|X|}
\hline
\textbf{Parameter} & \textbf{Symbol} & \textbf{Value} & \textbf{Units} & \textbf{Description} \\
\hline
Radius & \( R \) & \( 1.4\times 10^{12} \) & cm & Emission region radius \\
Magnetic field & \( B \) & \( 1.3\times 10^5 \) & G & Comoving magnetic field strength \\
\hline
X-ray min energy &  \( \varepsilon_{X,\rm min} \) & \( 100 \) & eV & Minimum energy of X-ray field \\
X-ray max energy &  \( \varepsilon_{X,\rm max} \) & \( 100 \) & keV & Maximum energy of X-ray field \\
X-ray luminosity & \( L_{X} \) & \( 5\times 10^{43} \) & erg/s & Luminosity of the X-ray field \\
\hline
Proton min Lorentz factor & \( \gamma_{p,\min} \) & \( 1 \) & – & Minimum Lorentz factor of protons \\
Proton break Lorentz factor & \( \gamma_{p,\rm br} \) & \( 2.7\times 10^4 \) & – & Break Lorentz factor of protons \\
Proton max Lorentz factor & \( \gamma_{p,\max} \) & \( 10^8 \) & – & Maximum Lorentz factor of protons \\
First proton spectral index & \( s_p \) & 1.0 & – & Power-law index of proton injection, below break \\
Second proton spectral index & \( s_p \) & 3.0 & – & Power-law index of proton injection, above break \\
Proton injection luminosity & \( L_p \) & \( 10^{43} \) & erg/s & Total injected proton luminosity \\
\hline
\end{tabularx}
\caption{Model parameters for the emission region and particle distributions in the AGN corona benchmark.}
\label{tab:corona_parameters}
\end{table}

\begin{figure*}
    \includegraphics[width=\textwidth]{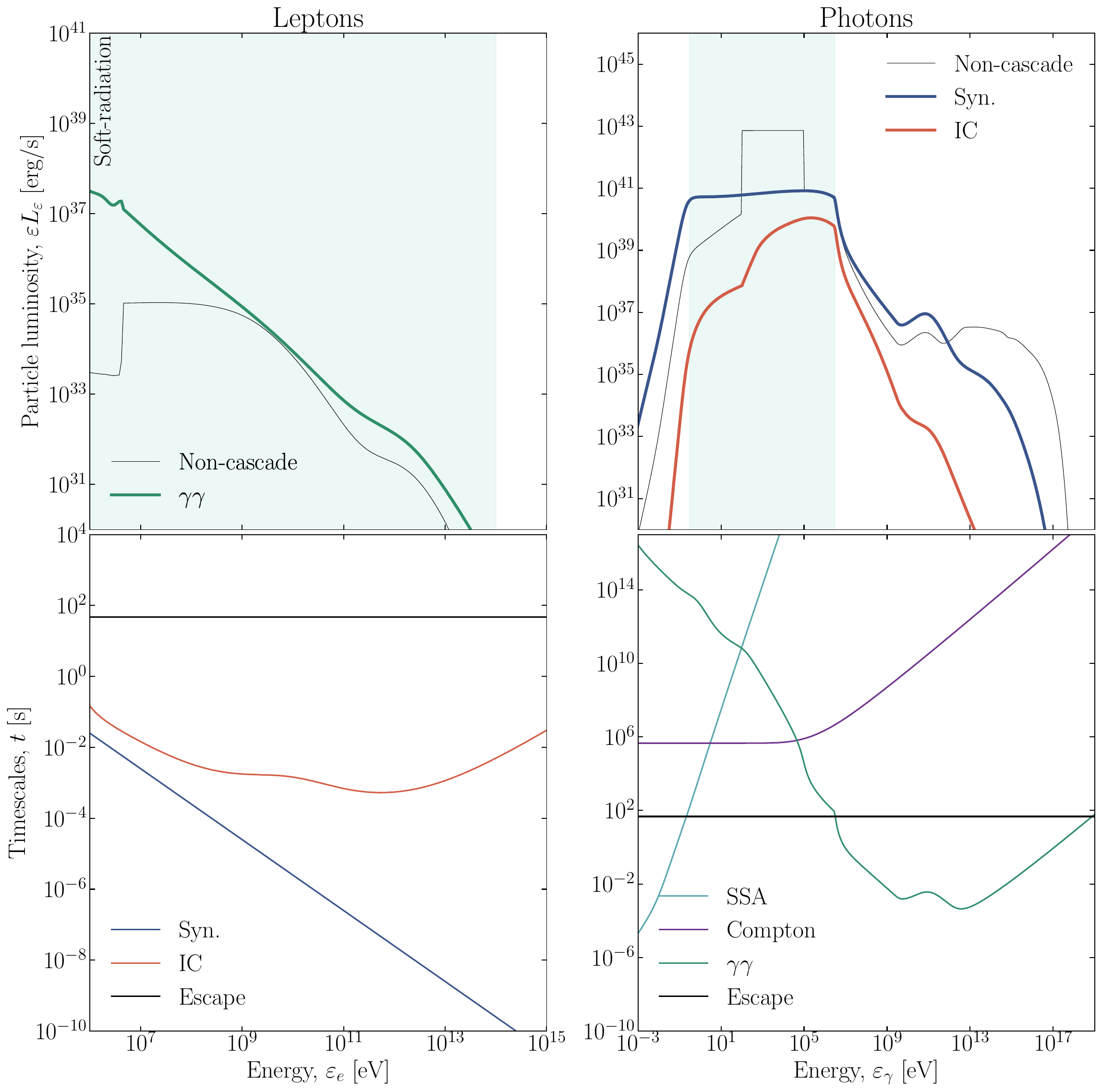}
    \caption{\textbf{Electromagnetic cascade in an AGN-corona-inspired hadronic model.}  
    Lepton (left) and photon (right) spectra in the coronal benchmark from Table~\ref{tab:corona_parameters}. Top: luminosity; bottom: interaction and escape timescales (synchrotron radiation, IC scattering, SSA, Compton scattering, and $\gamma\gamma$ pair production). Pair-production–driven components are colored; non-cascade particles are black. Background shading marks cascade regimes for pairs, and for photons the synchrotron emission from leptons in the corresponding regime. Most pairs are injected at $\varepsilon_e \simeq \gamma_{p,\rm br} m_p$ and promptly enter the soft-radiation cascade, with synchrotron-dominated losses, yielding a flat $s_\gamma=2$ spectrum. The non-cascade pairs come mostly from BH, and is subdominant (the sharp BH drop reflects numerical-method matching in \texttt{AM$^3$} ).}    
\label{fig:corona}
\end{figure*}

Due to the larger $\gamma\gamma$ opacity, the predominant electromagnetic emission beyond the leptonic X-rays would be associated with the electromagnetic cascade triggered by the high-energy hadronic injection. Therefore, this is a paradigmatic case of interest for the theory we have developed in this work. To see how the cascade emission in this case fits into our general framework, we now simulate the radiative emission from the corona. The specific properties of non-thermal emission are quite sensitive to the acceleration mechanism of high-energy protons, which in turn depend on the radiative compactness of the corona. For a very compact corona, with typical size below ten gravitational radii, acceleration in magnetized reconnection layers is the preferred scenario, introduced in Refs.~\cite{Fiorillo:2023dts, Karavola:2024uui}. Instead, for a somewhat less compact corona, with size of the order of tens of gravitational radii, a slower acceleration in strongly magnetized turbulence is more likely~\cite{Fiorillo:2024akm} (see Ref.~\cite{Murase:2019vdl} for a treatment of gyroresonant acceleration in stochastic turbulence, which however seems energetically incompatible with the neutrino luminosity inferred by IceCube unless large deviations from quasi-linear theory are assumed~\cite{Fiorillo:2024akm}).
Here we focus on the former reconnection scenario, and use \texttt{AM$^3$}  to simulate the radiative evolution of the setup described in Table~\ref{tab:corona_parameters}. These parameters are motivated by the NGC~1068 scenario considered in Ref.~\cite{Fiorillo:2023dts}. Notice that in this case the proton spectrum is characteristically a broken power law, rather than a single power law. We show all the components which do not directly originate from $\gamma\gamma$ interactions as non-cascade components; these include the primary X-rays and the leptons injected by BH and by $p\gamma$ interactions. Generally, it is not possible to distinguish systematically between cascade and non-cascade components -- even leptons injected by BH and by $p\gamma$ will produce photons which in turn will pair-produce. However, under our assumption of linear cascade in which the high-energy gamma rays are a perturbation on top of the large background flux, this distinction is possible.

The pairs produced at very high energies by photohadronic interactions cascade down to low energies driven primarily by synchrotron radiation, as is visible from the timescale plot. Therefore, they rapidly settle in the soft-radiation regime with $s_e=3$. The corona is in fact so compact and has such a large magnetic field that the pairs persist in this state down to their rest-mass energy. In turn, the synchrotron radiation from the corona is extremely flat, with $s_\gamma=2$ across a wide energy range reaching down to the minimum energy, which is determined by SSA. Therefore, the corona is another case in which the synchrotron-dominated cascade emerges naturally. These results are completely in agreement with previous numerical studies of the cascade in coronal environments~\cite{Fiorillo:2023dts, Karavola:2024uui, Fiorillo:2025cgm}. In scenarios with stochastic acceleration~\cite{Murase:2019vdl,Fiorillo:2024akm,Fiorillo:2025ehn}, the predominant role of BH interactions may break the universality by introducing an injection range extended in energy over a wide range.

\subsection{Gamma-Ray Bursts}

Gamma-Ray Bursts (GRBs) are among the most luminous and energetic transients in the universe, emitting intense flashes of gamma rays over timescales ranging from milliseconds to minutes. Their prompt emission is characterized by a highly non-thermal spectrum, typically extending from keV to GeV energies, and is thought to originate from relativistic outflows powered by compact central engines. The dominant processes shaping this spectrum are leptonic in origin -- IC scattering and synchrotron radiation from a population of high-energy non-thermal leptons. On the other hand, lepto-hadronic models have also been proposed to explain high-energy features observed by Fermi-LAT~\cite{Asano:2012jr,Wang:2018xkp,Rudolph:2022ppp}. In these cases, the hadronic particles produce high-energy neutrinos and gamma rays, with the latter being attenuated by the $\gamma\gamma$ interaction with the dominant leptonic radiation, and being reprocessed at low energies via the electromagnetic cascade. As we will see, this cascade exhibits the universal behavior we have identified.

\begin{table}[h!]
\centering
\begin{tabular}{|l|c|c|c|l|}
\hline
\textbf{Parameter} & \textbf{Symbol} & \textbf{Value} & \textbf{Units} & \textbf{Description} \\
\hline
Radius & \( R \) & \( 10^{13} \) & cm & Emission region radius \\
Magnetic field & \( B \) & \( 2 \times 10^3 \) & G & Comoving magnetic field strength \\
\hline
Electron min Lorentz factor & \( \gamma_{e,\min} \) & \( 10^4 \) & – & Minimum Lorentz factor of electrons \\
Electron max Lorentz factor & \( \gamma_{e,\max} \) & \( 5 \times 10^6 \) & – & Maximum Lorentz factor of electrons \\
Electron spectral index & \( s_e \) & 2.8 & – & Power-law index of electron injection \\
Electron injection luminosity & \( L_e \) & \( 10^{42} \) & erg/s & Total injected electron luminosity \\
\hline
Proton min Lorentz factor & \( \gamma_{p,\min} \) & \( 10^4 \) & – & Minimum Lorentz factor of protons \\
Proton max Lorentz factor & \( \gamma_{p,\max} \) & \( 5 \times 10^7 \) & – & Maximum Lorentz factor of protons \\
Proton spectral index & \( s_p \) & 2.0 & – & Power-law index of proton injection \\
Proton injection luminosity & \( L_p \) & \( 10^{42} \) & erg/s & Total injected proton luminosity \\
\hline
\end{tabular}
\caption{Model parameters for the emission region and particle distributions in the GRB benchmark. All the parameters are defined within the comoving frame of the dissipation region.}
\label{tab:grb_parameters}
\end{table}

\begin{figure*}
    \includegraphics[width=\textwidth]{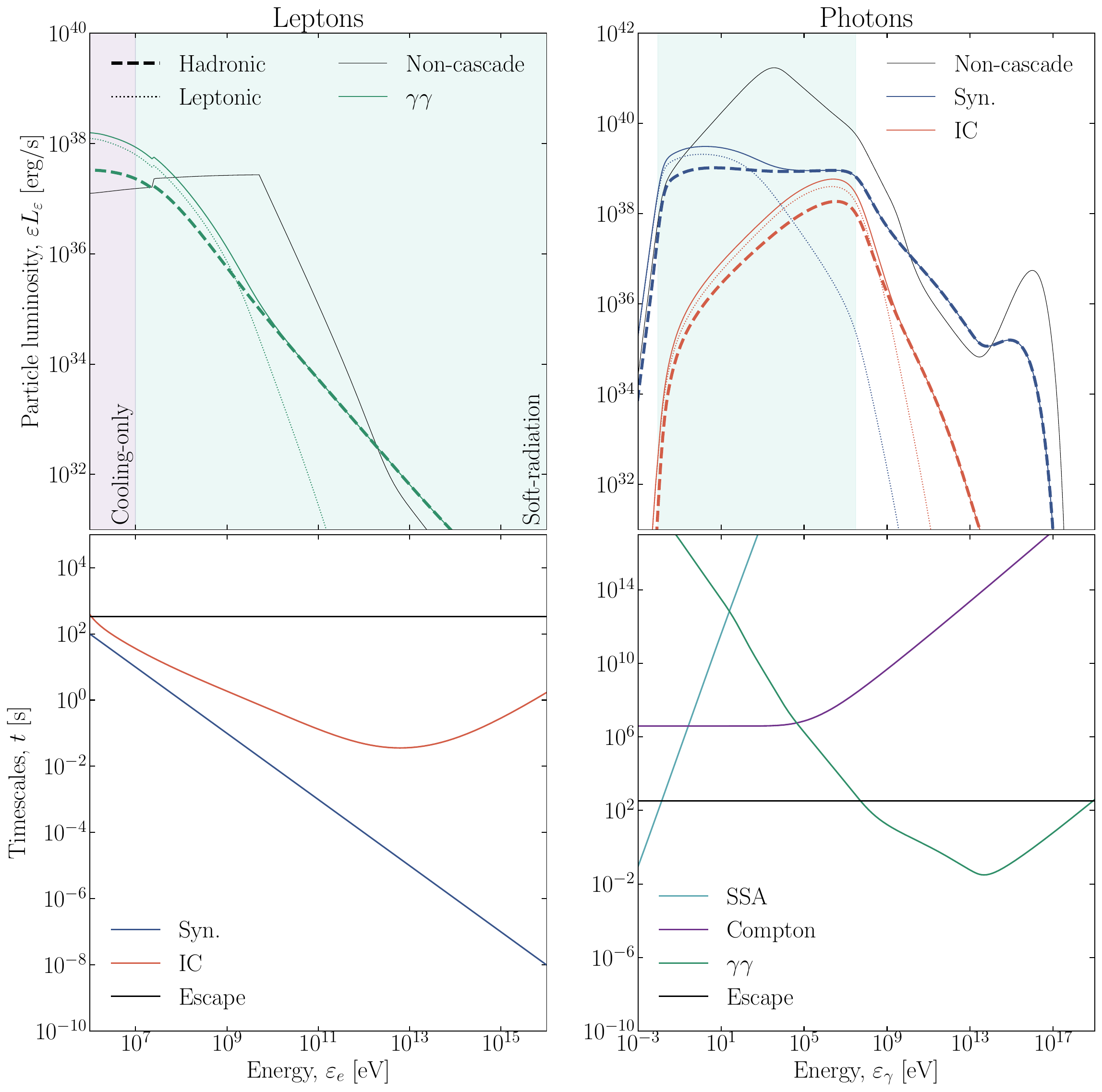}
    \caption{\textbf{Electromagnetic cascade in a GRB-inspired lepto-hadronic model.}  
Same as Fig.~\ref{fig:corona} for a GRB benchmark. Leptonic and hadronic contributions are distinguished by line style. Background shading indicates the cascade regimes for the pairs: a high-energy soft-radiation regime (set by pair production and synchrotron cooling) and a low-energy cooling-only regime without pair injection. For photons, the background shading highlights the energy range where photons are produced by leptons in the corresponding cascade regime. Particles which do not come from $\gamma\gamma$ (non-cascade) are shown in black.
}\label{fig:grb}
\end{figure*}

To show this, we use again the numerical code \texttt{AM$^3$} , and prepare a setup analogous to the lepto-hadronic models in Refs.~\cite{Rudolph:2022ppp,Rudolph:2022dky}. The parameters used for this setup are summarized in Table~\ref{tab:grb_parameters}. The simulation is carried out entirely in the comoving frame of the dissipation region, so that we do not have to specify the Doppler boosting factor of the region. Fig.~\ref{fig:grb} shows the resulting electromagnetic emission, as well as the characteristic energy-dependent timescales for all energy-loss and escape processes of pairs and photons. For the pair-induced emission (shown in color in the figure), we clearly differentiate in these figures between the purely leptonic emission (dotted, thin), which is obtained by simulating the dissipation region with $L_p=0$, and the hadronic component (dashed, thick). The primary emission, shown in black, shows the typical prompt gamma-ray emission peaking in the 100~keV-1~MeV range, produced by the leptons injected above $10^{10}\,\mathrm{MeV}$. The minimum Lorentz factor for the injected leptons is also the cause for the abrupt break in the primary lepton spectrum, below which the pairs enter the fast-cooling regime, as in the cases of Ref.~\cite{Rudolph:2022ppp}.

The hadronic cascade follows precisely the dynamics of the synchrotron-dominated regime we have identified. At high energies, the pairs are continuously produced by the photons, which in turn are replenished by the soft synchrotron radiation. The corresponding soft-radiation cascade has a spectral index $s_e=3$, and produces radiation with $s_\gamma=2$. At low energies, photons manage to escape, leading to a narrow interval of cooling-only cascade with $s_e=2$; these pairs ultimately produce photons in the low-energy range where they are synchrotron-self-absorbed, so the corresponding power law with $s_\gamma=3/2$ is not visible. The lepto-hadronic, synchrotron-dominated regime considered in Ref.~\cite{Rudolph:2022ppp} always shows a qualitatively similar structure to this case, which is described by the cascade regime derived here. As for the leptonic component, as anticipated in Sec.~\ref{sec:limitations}, its injection extended over a wide energy range masks the universality of the cascade and leads to a spectrum which depends on the specifics of the injected particles. In particular, the thin dotted green line in Fig.~\ref{fig:grb} settles into a power law much softer than $s_\gamma=3$, following the injection of the primary leptons. In Appendix~\ref{app:ic_dominated_grb}, we also consider a GRB-inspired benchmark in which the cascade is instead IC-dominated; as discussed in Sec.~\ref{sec:limitations}, such cases generally fail to reproduce a universal cascade due to the non-thermal nature of the target photons for IC losses.

\subsection{Active Galactic Nuclei (AGN) blazars}

Blazars are a subclass of AGN powered by accreting supermassive black holes, with relativistic jets closely aligned with our line of sight. This orientation leads to strong Doppler boosting of the jet emission, making blazars some of the most luminous persistent sources in the gamma-ray sky. Their spectral energy distributions (SEDs) exhibit two broad non-thermal components: a low-energy bump attributed to synchrotron radiation from relativistic electrons, and a high-energy bump typically extending from X-rays to TeV gamma rays. Energy dissipation likely occurs in compact regions along the jet, at parsec or sub-parsec scales, where shocks, magnetic reconnection, or turbulence can accelerate particles to ultra-relativistic energies. 

While purely leptonic models -- in which the high-energy component arises from IC scattering by electrons -- can explain many observed features in the electromagnetic SED, they do not lead to neutrino emission. Thus, the recent association of neutrinos with blazars, particularly after the case of TXS~0506+056, has motivated the development of lepto-hadronic scenarios, with the hadronic component showing up primarily in neutrino emission and in the electromagnetic cascade.

\begin{table}[h!]
\centering
\begin{tabular}{|l|c|c|c|l|}
\hline
\textbf{Parameter} & \textbf{Symbol} & \textbf{Value} & \textbf{Units} & \textbf{Description} \\
\hline
Radius & \( R \) & \( 10^{17} \) & cm & Emission region radius \\
Magnetic field & \( B \) & \( 0.1 \) & G & Comoving magnetic field strength \\
\hline
Electron min Lorentz factor & \( \gamma_{e,\min} \) & \( 1 \) & – & Minimum Lorentz factor of electrons \\
Electron max Lorentz factor & \( \gamma_{e,\max} \) & \( 5 \times 10^6 \) & – & Maximum Lorentz factor of electrons \\
Electron spectral index & \( s_e \) & 2 & – & Power-law index of electron injection \\
Electron injection luminosity & \( L_e \) & \( 10^{42} \) & erg/s & Total injected electron luminosity \\
\hline
Proton min Lorentz factor & \( \gamma_{p,\min} \) & \( 1 \) & – & Minimum Lorentz factor of protons \\
Proton max Lorentz factor & \( \gamma_{p,\max} \) & \( 10^{11} \) & – & Maximum Lorentz factor of protons \\
Proton spectral index & \( s_p \) & 2.0 & – & Power-law index of proton injection \\
Proton injection luminosity & \( L_p \) & \( 10^{42} \) & erg/s & Total injected proton luminosity \\
\hline
\end{tabular}
\caption{Model parameters for the emission region and particle distributions in the blazar benchmark. All the parameters are defined within the comoving frame of the dissipation region.}
\label{tab:blazar_parameters}
\end{table}

\begin{figure*}
    \includegraphics[width=\textwidth]{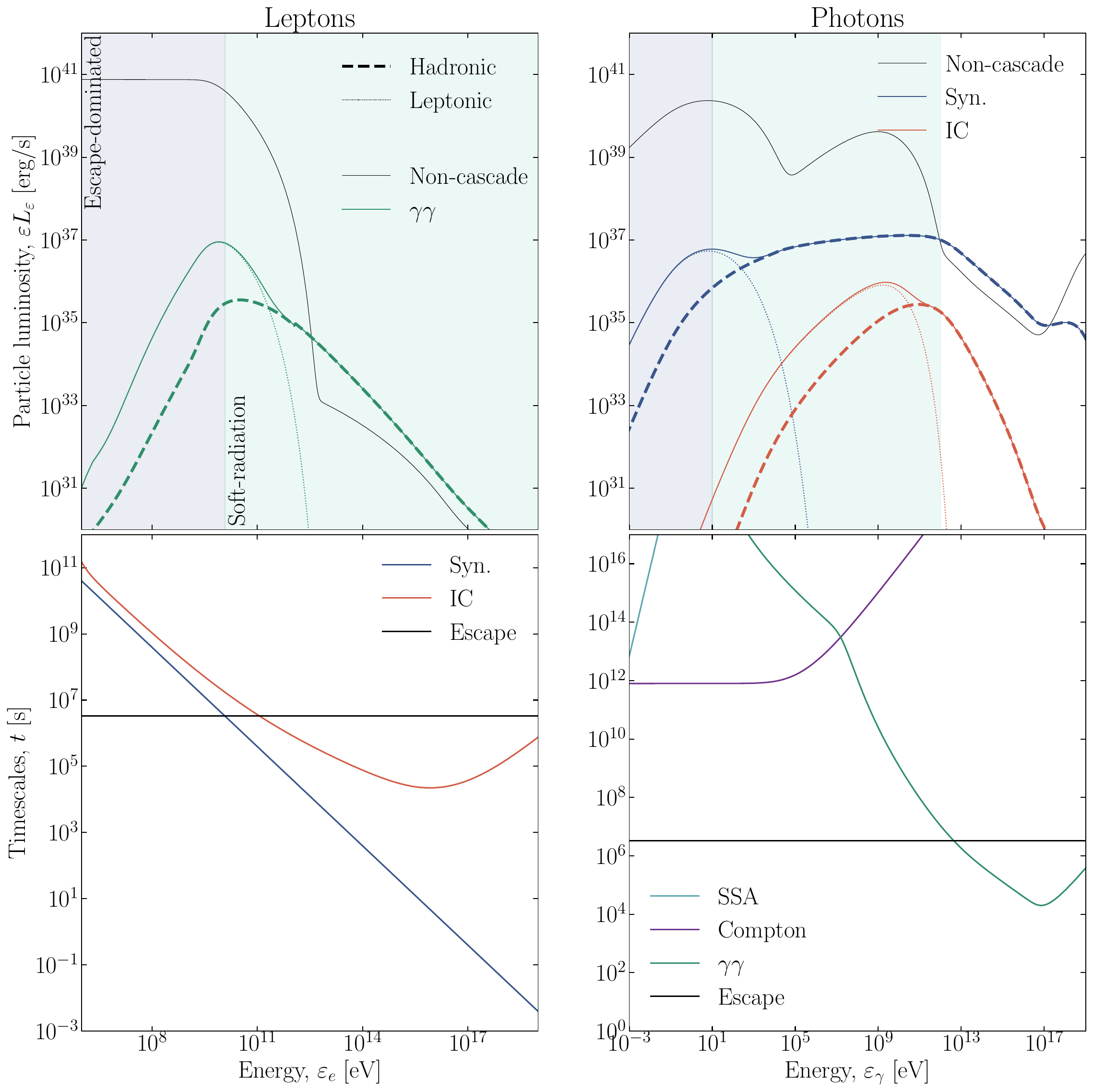}
    \caption{\textbf{Electromagnetic cascade in a blazar-inspired lepto-hadronic model.}  Same as Fig.~\ref{fig:grb} for a blazar benchmark. The numerical values of the parameters are collected in Table~\ref{tab:blazar_parameters}. Particles which do not come from $\gamma\gamma$ (non-cascade) are shown in black. The pairs transition from the soft-radiation cascade at high energies to the escape-dominated regime at low energies, producing the characteristic cascade with $s_\gamma=2$.}\label{fig:blazar}
\end{figure*}

It is impossible to capture the variety of potential evolution of the radiative signature of lepto-hadronic blazar models, that cover a wide range of parameters. The appearance of the universal cascade we have identified is far from universal; on the other hand, if the dissipation region is sufficiently compact, and the BH injection is not the dominant production channel for secondaries, it can still emerge. We show one such example, using the benchmark parameter summarized in Table~\ref{tab:blazar_parameters}. Fig.~\ref{fig:blazar} shows the resulting pair and electromagnetic emission. The non-cascade emission shows the characteristic two-humped structure, with the first hump due to synchrotron radiation and the second one due to IC radiation of the pairs interacting with the first hump. 

The hadronic cascade, instead, reproduces the main features we have identified for the synchrotron-dominated cascade, which indeed is the case realized as visible from the timescale plot. Thus, pairs at high energies are in the soft-cascade regime, with the spectral index $s_e=3$, down to a low-energy threshold below which pairs are escape-dominated. Correspondingly, the radiation exhibits the typical $s_\gamma=2$ cascade that we have observed in the GRB case as well, suppressed at low energies by the escape-dominated regime. Also in this case, as in the GRB one, the leptonic pair-induced cascade does not obey the universal cascade form, due to the very different injection spectrum which extends throughout the range where the cascade develops.

Clearly a single example is not meant to show that the hadronic cascade in blazars is always of the universal, synchrotron-dominated nature. In fact, in the majority of lepto-hadronic blazar models considered in the literature, the cascade does \textit{not} exhibit the universal spectrum obtained in this work. It is instructive to consider a few cases, to understand more clearly what assumptions are broken and lead to a non-universal behavior. The first assumption that can easily be broken is that of a fully developed cascade, which requires the dissipation region to be optically thick to high-energy gamma rays and leptons. For example, for most models proposed to explain the neutrino association with TXS~0506+056, the high-energy gamma rays are only partially absorbed, see e.g. Refs.~\cite{Gao:2018mnu,Petropoulou:2019zqp}. Thus, the resulting cascade spectrum is not a flat power law, but rather a collection of bumps, as visible in Fig.~3 of Ref.~\cite{Gao:2018mnu}. The universal cascade might also be hindered by a predominance of BH processes, leading to an injection of pairs over a wide energy range; many examples in the literature of this type, see e.g. the benchmark in Fig.~2 of Ref.~\cite{Klinger:2023zzv}, and the systematic study of Ref.~\cite{2024Univ...10..392C}. 

\subsection{Tidal disruption events}

Tidal disruption events (TDEs) occur when a star is torn apart by the tidal forces of a supermassive black hole, producing a luminous flare powered by the accretion of stellar debris. While a small subset of TDEs exhibit relativistic jets and gamma-ray emission, the majority are non-jetted and characterized by softer, thermal-like emission, typically in the infrared (IR), optical/ultra-violet (OUV), and X-ray bands. In these systems, particle acceleration may still occur within the debris streams, the accretion disk, and outflows, but the absence of strong non-thermal leptonic signatures in neutrino-emitting TDEs and TDE candidates suggests suppressed or subdominant leptonic loading. From the theoretical point of view, in diffusive shock acceleration there is grounds to expect a baryonic loading quite larger than the leptonic one; see the discussion and references in Ref.~\cite{Yuan:2023cmd}. Consequently, we will assume that only protons are injected as non-thermal particles, particularly when motivated by the potential for high-energy neutrino production. These hadronic scenarios for non-jetted TDEs have recently gained traction following reports of four temporal coincidences between IceCube events and flaring TDEs~\cite{Winter:2020ptf,Reusch:2021ztx,Winter:2022fpf,Yuan:2024foi}.

\begin{table}[h!]
\centering
\begin{tabular}{|l|c|c|c|l|}
\hline
\textbf{Parameter} & \textbf{Symbol} & \textbf{Value} & \textbf{Units} & \textbf{Description} \\
\hline
Radius & \( R \) & \( 5\times 10^{17} \) & cm & Emission region radius \\
Magnetic field & \( B \) & \( 0.1 \) & G & Comoving magnetic field strength \\
\hline
IR temperature &  \( T_{\rm IR} \) & \( 0.16 \) & eV & Temperature of the IR field \\
IR luminosity & \( L_{\rm IR} \) & \( 10^{45} \) & erg/s & Luminosity of the IR field \\
OUV temperature &  \( T_{\rm OUV} \) & \( 1.3 \) & eV & Temperature of the OUV field \\
OUV luminosity & \( L_{\rm OUV} \) & \( 10^{45} \) & erg/s & Luminosity of the OUV field \\
\hline
Proton min Lorentz factor & \( \gamma_{p,\min} \) & \( 1 \) & – & Minimum Lorentz factor of protons \\
Proton max Lorentz factor & \( \gamma_{p,\max} \) & \( 5.3\times 10^9 \) & – & Maximum Lorentz factor of protons \\
Proton spectral index & \( s_p \) & 2.0 & – & Power-law index of proton injection \\
Proton injection luminosity & \( L_p \) & \( 10^{42} \) & erg/s & Total injected proton luminosity \\
\hline
\end{tabular}
\caption{Model parameters for the emission region and particle distributions in the TDE benchmark.}
\label{tab:tde_parameters}
\end{table}

\begin{figure*}
    \includegraphics[width=\textwidth]{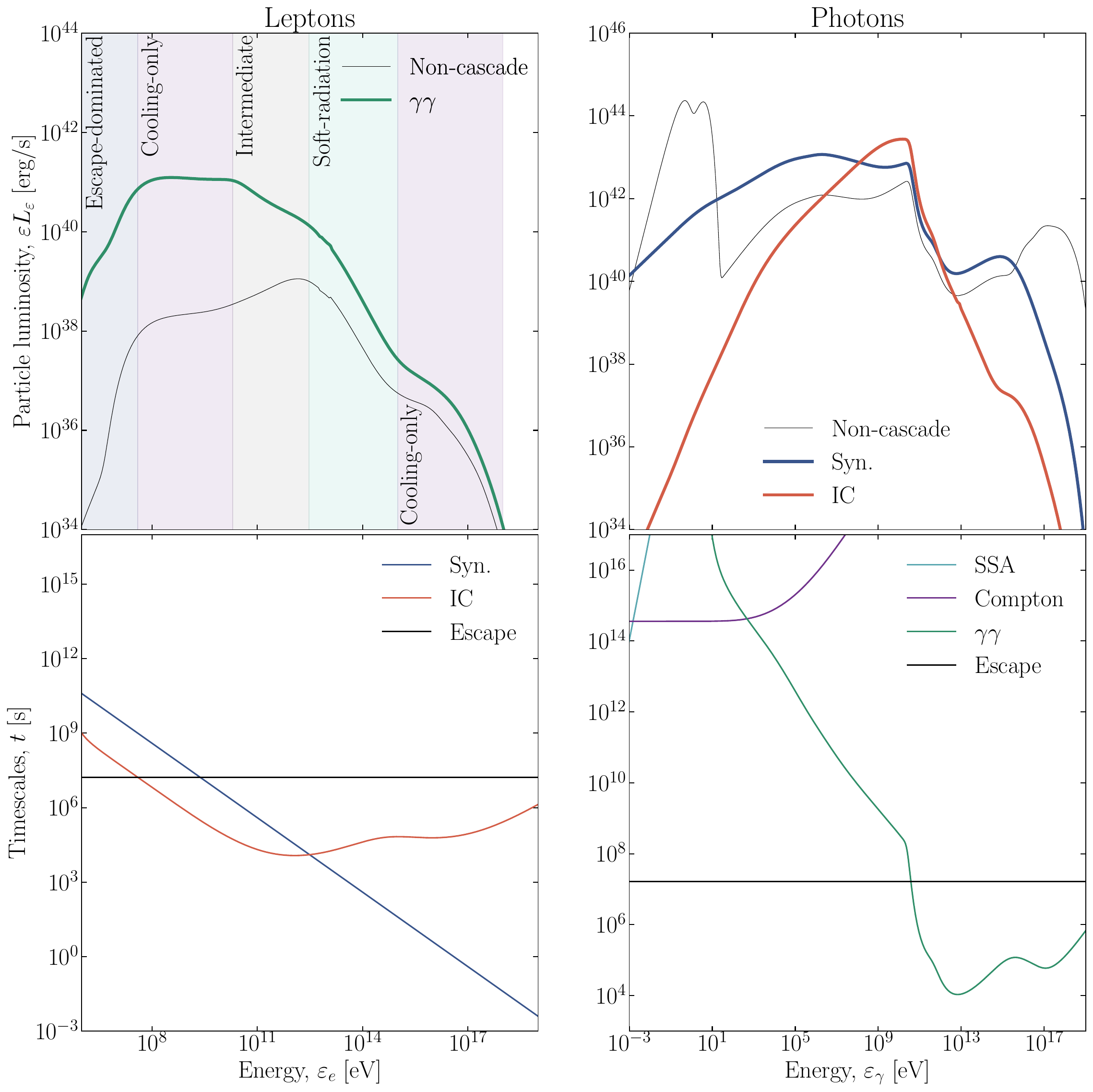}
    \caption{\textbf{Electromagnetic cascade in a TDE-inspired hadronic model.}  
Same as Fig.~\ref{fig:corona}, but for the TDE benchmark (parameters in Table~\ref{tab:tde_parameters}).  
For photons, background shading is not used to indicate different cascade regimes, as IC and synchrotron components -- comparable in this case -- dominate in different regions. Particles which do not come from $\gamma\gamma$ (non-cascade) are shown in black.  
At high energies, pair losses are synchrotron-dominated, and pairs transition from the cooling-only to the soft-radiation regime.  
Below $\sim\!10^{12}\,\mathrm{eV}$, IC scattering becomes the dominant cooling channel, placing pairs in an intermediate regime.  
Below the pair-production threshold, they re-enter a cooling-only regime, and at the lowest energies, they eventually reach the escape-dominated regime.}\label{fig:tde}
\end{figure*}

Under this assumption, electromagnetic cascades initiated by photohadronic interactions of relativistic protons become the primary source of high-energy photon emission, which motivates an attempt at qualitatively understanding them. Thus, we provide in this section a TDE-like benchmark and discuss its radiative cascade properties. The primary features of this benchmark is the absence of leptonic emission; the target for $\gamma\gamma$ attenuation is external, dominated by the thermal IR and OUV radiation field. We consider a simplified yet representative TDE cascade model in which accelerated protons are injected into a spherically symmetric radiation zone of radius $R \sim 10^{16}$–$10^{18}\rm cm$, permeated by a magnetic field of strength $B \sim 0.1~\rm G$. In this region, thermal IR and OUV photons are isotropized and serve as targets for photohadronic ($p\gamma$) interactions. These interactions produce high-energy $\gamma$-rays through $\pi^0$ decays, initiating subsequent electromagnetic cascades primarily governed by $\gamma\gamma$ annihilations between the $\pi^0$-decay $\gamma$-rays and the dense target photon fields. The key parameters adopted for this simulation (consistent with those used in Ref. \cite{Yuan:2023cmd}) are summarized in Table~\ref{tab:tde_parameters}. Without loss of generality for the cascade modeling, we fix the luminosities of the injected protons and target photons to obtain the steady-state spectra, while in reality, these quantities may be time-dependent.

Fig.~\ref{fig:tde} shows the corresponding lepton and photon spectrum. The cascade spectrum is significantly more intricate than previous examples. This is due to the coexistence of IC and synchrotron losses, which are comparable and dominate across different energy ranges. At large energies, where the gamma rays are injected, synchrotron radiation is the dominant cooling channel. Above $10^{15}\, \mathrm{eV}$, synchrotron photons cannot be produced because their typical frequency is too low, and therefore no pairs are injected, leading to a cooling-only regime. Instead, between $10^{13}\,\mathrm{eV}\lesssim \varepsilon_e \lesssim 10^{15}\,\mathrm{eV}$, the pairs exhibit the $s_e=3$ index characteristic of the soft-radiation cascade. 
At lower energies, IC scattering becomes the dominant energy loss mechanism, so the pairs are in an equal-reproduction regime, which however is contaminated by the competing synchrotron losses so that no characteristic spectral shape can be identified, down to the energy marked by the dotted line, around $10^{11}\,\mathrm{eV}$. At lower energies, pair production is inefficient, so we enter the cooling-only regime, with the characteristic $s_e=2$. Finally, at sufficiently low energies, below about $\varepsilon_e\lesssim 10^8\,\mathrm{eV}$, there is the escape-dominated regime.

In turn, the two components of the electromagnetic cascade, synchrotron and IC radiation, also exhibit their characteristic features in the respective energy ranges. The IC component dominates at high-energies, where it has a narrow energy range in which it is roughly flat, as expected for a bichromatic target spectrum -- which is a good approximation for the two thermal bumps of the assumed radiation field -- and at lower energies drops with $s_\gamma=3/2$. Instead, the synchrotron component exhibits a wide range in which it is roughly flat, originating from the pairs in the soft-radiation regime. These generalized conclusions are in good agreement with the cascade spectra obtained from time-dependent modeling of neutrino-emitting TDEs \cite{Yuan:2023cmd,Yuan:2024foi} and TDE-like AGN flares \citep{Yuan:2025zwe}. Thus, overall the TDE case reveals individual traits of the universal cascade we had previously identified, although it is as a whole more complex in its details because of the coexistence of IC and synchrotron losses, which dominate in different energy ranges.

\section{Discussion}\label{sec:discussion}

Understanding electromagnetic cascades from hadronic injection is increasingly crucial, as they may represent the only observable electromagnetic (non-neutrino) signature of hadronic acceleration. This is particularly evident in the few cases where neutrinos correlate with known astrophysical sources. For instance, during the 2017 flare of TXS~0506+056, X-rays provided the dominant constraint on neutrino luminosity~\cite{Keivani:2018rnh, Murase:2018iyl, Cerruti:2018tmc, Gao:2018mnu}, since reprocessed gamma rays could not exceed the observed X-ray flux. Constraints are even stronger for the 2014--15 neutrino excess, which lacked any gamma-ray counterpart~\cite{Keivani:2018rnh, Rodrigues:2018tku}. Even scenarios involving coronal neutrino production face severe limits~\cite{Fiorillo:2025cgm}, disfavoring them for this source. In the context of gamma-ray opaque sources, like the cores of AGN, the cascade is the primary electromagnetic signature of hadronic acceleration.

These cases motivate the need for reliable, qualitative estimates of the reprocessed cascade emission -- its magnitude and spectral range -- complementing numerical simulations that are often computationally expensive and not easily generalizable. Our results allow for such a qualitative understanding; we highlight the potential emergence of a universal cascade spectrum, generalizing the results of Berezinsky for extragalactic cascade, and point out the conditions in which this spectrum can be realized. The consequences of this result span different astrophysical sources.

For blazars, the speculation that the cascade from TXS~0506+056 could follow a $s_\gamma=2$ power law between 30~TeV and 3~PeV~\cite{Halzen:2018iak} appears inconsistent with realistic source conditions. As our blazar-inspired benchmark illustrates (Fig.~\ref{fig:blazar}), a $s_\gamma=2$ spectrum may arise if synchrotron losses dominate, but the spectral range is set by physical thresholds: the synchrotron frequency of the lowest-energy pairs and the minimum energy for pair production. Even worse, generally speaking the condition for the universal spectrum with $s_\gamma=2$ are \textit{not} met; BH injection and partial attenuation of gamma rays are rather common in lepto-hadronic blazar models, which therefore exhibit a much more model-dependent electromagnetic emission. The conditions for a universal, synchrotron-dominated cascade are instead met more commonly in lepto-hadronic models of GRBs. For this case, our framework offers a direct explanation of the spectra numerically found in multiple works on the subject~\cite{Petropoulou:2014awa,Rudolph:2022ppp}.

The cascade structure becomes even more critical in systems such as TDEs and AGN coronae, where hadronic processes may dominate the high-energy photon emission. In TDEs, while we have not performed a full parameter scan (e.g., as in~\cite{Yuan:2023cmd}), we have shown that the cascade spectra can be understood using the general framework developed here. For AGN coronae, synchrotron-dominated cascades seem almost unavoidable under the magnetic reconnection scenario, due to the natural scaling between magnetic fields and X-ray energy density inferred from PIC simulations~\cite{Sironi:2015eoa}; in model-independent scenarios, the role of synchrotron-dominated cascades has also been emphasized in Refs.~\cite{Das:2024vug,Murase:2022dog}. This conclusion may change depending on the assumed acceleration mechanism. For instance, if strongly magnetized turbulence is responsible for particle energization, BH losses may become dominant over photohadronic interactions~\cite{Fiorillo:2024akm}, potentially allowing the cascade spectrum to discriminate the proton acceleration mechanism.

Let us also briefly discuss the relation of our work with previous analytical and numerical studies of electromagnetic cascades. Besides extragalactic cascades~\cite{Berezinsky:1975zz,Berezinsky:2016feh}, these have been proposed to explain AGN corona X-rays via non-linear pair cascades, where the main target photons are themselves produced by the cascade~\cite{Svensson:1987nlx}, making their dynamics distinct from the linear, hadronic cascades studied here. Similarly, Ref.~\cite{1997MNRAS.285...69B} considered pair cascades in blazars driven by curvature radiation and synchrotron pair production, again physically different from our setting. Instead, cascades driven by IC scattering internally to compact astrophysical sources have been discussed in the past; see, e.g., the seminal numerical discussion of Ref.~\cite{1987ApJ...319..643L} and especially Ref.~\cite{1988ApJ...335..786Z}. The results of this study are essentially in agreement with the Berezinsky theory of extragalactic cascades, although a connection between the two cases does not seem to have been acknowledged. In particular, Ref.~\cite{1988ApJ...335..786Z} predicts the $s_\gamma=-3/2$ low-energy behavior for the cascade from a monochromatic or blackbody target photon field. We should also stress that our results holds only for linear cascades, with no self-interaction among cascade particles. Non-linear cascades can exhibit a much more complex phenomenology with intrinsic temporal features, including oscillatory behavior; see, e.g., Refs.~\cite{1992Natur.360..135K,2012MNRAS.421.2325P,2020MNRAS.495.2458M}. Generally, in lepto-hadronic models in which the hadronic component is subdominant compared to the leptonic one, non-linear effects are expected to be small.

The broader significance of our results lies in the growing realization that a large fraction of extragalactic neutrino sources are likely to be opaque to gamma rays. AGN coronae may represent a substantial or even dominant class of such gamma-ray-dark sources. In these environments, the cascade emission is the primary electromagnetic signature -- both for individual sources~\cite{Murase:2019vdl, Fiorillo:2023dts, Karavola:2024uui} and for the diffuse background~\cite{Fang:2022trf}. The assumption in~\cite{Fang:2022trf} that such cascades resemble the Berezinsky type is invalid due to synchrotron losses, which are in fact dominant in magnetically powered environments. This dominance is not specific to the coronal model: any source powered by magnetic dissipation will feature comparable magnetic and radiative energy densities, placing synchrotron losses at the center of the cascade dynamics. From this viewpoint, the generalized cascade theory developed here -- incorporating both IC- and synchrotron-dominated regimes -- is essential for interpreting high-energy emission from gamma-ray-dark neutrino sources. 

\section{Summary}\label{sec:summary}

In this work, we developed a generalized theory of electromagnetic cascades—defined as the reprocessing of high-energy gamma rays via pair production and pair radiation—that extends the classic Berezinsky treatment of IC-dominated cascades~\cite{Berezinsky:1975zz,Berezinsky:2016feh} to include cases where synchrotron losses dominate. The latter appears to be a generic feature of \textit{internal} cascades produced within astrophysical sources.

Our main results are:
\begin{itemize}
    \item For IC-dominated cascades with (bi)monochromatic targets (e.g., thermal bumps), we recover a Berezinsky-like broken power-law spectrum ($s_\gamma = 3/2 \to 2$). This occurs partially in TDEs, though synchrotron losses are also significant, resulting in hybrid behavior (see Fig.~\ref{fig:tde});
    \item For non-thermal targets extending as power laws ($0 < s_t < 2$), monochromatic approximations fail and cascade universality is lost: the pair and photon spectra depend sensitively on the target. An example is shown for GRBs in Appendix~\ref{app:ic_dominated_grb}, where the IC component follows a power law, but not the canonical $s_\gamma = 3/2$ (Fig.~\ref{fig:grb_com});
    \item For synchrotron-dominated cascades (treated here for the first time), the photon spectrum typically is a broken power law, with an extended high-energy range with $s_\gamma=2$, bounded below by the synchrotron frequency of the lowest-energy pairs. At lower energies, the spectrum breaks into $s_\gamma=3/2$. The break is caused by the corresponding transition in the pair spectrum from a high-energy soft-radiation regime to a low-energy cooling-only regime in which $\gamma\gamma$ pair production is interrupted. On the other hand, if the source is so compact that $\gamma\gamma$ pair production is efficient even at MeV energies, comparable with the electron mass, the break disappears and the photon spectrum has $s_\gamma=2$ down to very low energies, where it can be suppressed either by SSA or by the escape of the radiating pairs. 
    This regime emerges across all benchmarks—GRBs (Fig.~\ref{fig:grb}), blazars (Fig.~\ref{fig:blazar}), TDEs (Fig.~\ref{fig:tde}), and AGN coronae (Fig.~\ref{fig:corona})—when magnetic energy density is sufficiently large. We have shown it to be surprisingly universal, and in general independent of the shape of the target photon spectrum.
\end{itemize}

Hadronic cascades typically settle into one or a combination of these regimes, depending on the relative importance of IC and synchrotron losses.  A possible exception is when Bethe-Heitler (BH) losses dominate, injecting pairs across a broad energy range and potentially spoiling the cascade-down assumption. Since the process remains linear, a superposition approach could still apply, provided an analytical BH injection spectrum is available (e.g.,~\cite{Karavola:2024uog}); we leave this for future work.

Overall, we find that such hadronic cascades in the synchrotron-dominated regime are quite present in the literature, as they naturally appear in compact AGN coronae and lepto-hadronic models of GRBs. In blazars, while certain very compact setups may also lead to a similar spectral shape, lepto-hadronic models often produce a different, model-dependent cascade, due to the dominance of BH processes and to conditions of partial gamma-ray absorption. AGN coronae may also exhibit BH injection which alters the nature of the cascade, depending on the compactness of the acceleration zone; thus discriminating between the universal cascade identified here and a non-universal cascade affected by BH processes may offer an opportunity for inferring the compactness of the radiation zone. With the aid of the framework developed here, one can now clarify for any specific astrophysical environment whether the conditions are met for the appearance of a universal cascade, and directly relate its properties to the geometry and energetics of the source.

\section*{Acknowledgments}

We thank Mahmoud Alawashra and Kohta Murase for useful comments on this manuscript. The research project has been funded by the ``Program for the Promotion of Exchanges and Scientific Collaboration between Greece and Germany IKYDA--DAAD'' 2024 (IKY project ID 309; DAAD project ID: 57729829). 
D.F.G.F. is supported by the Alexander von Humboldt Foundation (Germany).  M.P. acknowledges support from the Hellenic Foundation for Research and Innovation (H.F.R.I.) under the ``2nd call for H.F.R.I. Research Projects to support Faculty members and Researchers" through the project UNTRAPHOB (Project ID 3013).

\appendix

\section{Steady-state pairs in soft-radiation cascade}\label{app:soft_radiation_cascade}

In this section, we prove directly from the kinetic equations the properties of the soft-radiation regime of the pair cascade. We focus on the synchrotron-dominated regime, for which the dominant pair energy losses are given by Eq.~\ref{eq:syn_loss}. The soft-radiation regime also appears in the IC-dominated case when the dominant energy loss of pairs is IC scattering in the Thomson regime off a low-energy target photon field with typical energies $\varepsilon_{t, \rm min}$, while pair production is still active thanks to the $\gamma\gamma$ scattering off higher-energy target photons at $\varepsilon_{t, \rm max}$. In this case, the dynamics is identical, since we still have the dominant energy loss term $b_{\rm IC}(\varepsilon_e)=(-d\varepsilon_e/dt)_{\rm IC}\propto \varepsilon_e^2$ and the typical frequency of the radiated IC photons much lower than the lepton energy $\varepsilon_e$, of the order of $\varepsilon_\gamma\simeq \varepsilon_{t, \rm min}(\varepsilon_e/m_e)^2$. We focus on the synchrotron case for definiteness.

In the soft-radiation cascade, the radiation of energy -- synchrotron in this case -- and pair emission and losses are balanced. 
The balance equation for leptons and photons has the same form as Eq.~\ref{eq:balance_inverse_compton}, except that IC losses must be replaced with synchrotron ones. Therefore, we have for $\varepsilon_e>\varepsilon_{e,\rm thr}$
\begin{eqnarray}\label{eq:balance_synchrotron}
    \frac{\partial n_e(\varepsilon_e)}{\partial t}&=&\frac{\partial}{\partial \varepsilon_e}\left[b_{\rm syn} (\varepsilon_e) n_e(\varepsilon_e)\right]+4n_\gamma(2\varepsilon_e) \Gamma_{\gamma\gamma}(2\varepsilon_e)=0,\\
    \frac{\partial n_\gamma(\varepsilon_\gamma)}{\partial t}&=&\frac{U_B \sigma_Tm_e}{2\omega_B}\sqrt{\frac{3}{4\omega_B \varepsilon_\gamma}}n_e\left[m_e\sqrt{\frac{3\varepsilon_\gamma}{4\omega_B}}\right]-n_\gamma(\varepsilon_\gamma)\Gamma_{\gamma\gamma}(\varepsilon_\gamma)=0,
\end{eqnarray}
where the synchrotron radiation is described by the delta-function approximation discussed above. Notice that we are neglecting a potential photon escape term; we assume photons are completely confined by pair production, so such a term would be inessential.

From the two equations, we can now eliminate $n_\gamma(\varepsilon_\gamma) \Gamma_{\gamma\gamma}(\varepsilon_\gamma)$ and obtain an equation for the pair number density only
\begin{equation}
    \frac{\partial}{\partial \varepsilon_e}\left[\varepsilon_e^2 n_e(\varepsilon_e)\right]+\frac{3m_e^3}{4\omega_B}\sqrt{\frac{3}{2\omega_B \varepsilon_e}}n_e\left[m_e \sqrt{\frac{3\varepsilon_e}{2\omega_B}}\right]=0.
\end{equation}
To understand the main properties of this solution, we now transform $n_e(\varepsilon_e)=\varepsilon_e^{-3} \Phi_e(\varepsilon_e)$, so as to obtain
\begin{equation}
    \frac{\partial}{\partial \varepsilon_e}\left[\frac{\Phi_e(\varepsilon_e)}{\varepsilon_e}\right]+\frac{1}{2\varepsilon_e^2}\Phi_e\left[m_e\sqrt{\frac{3\varepsilon_e}{2\omega_B}}\right]=0.
\end{equation}
All dimensional energy scales have dropped out of this equation; indeed, if we now write $\varepsilon_e=3m_e^2 e^x/2\omega_B$, and  we call $\tilde{\Phi}_e(x)=\left.\Phi_e(\varepsilon_e)\right|_{\varepsilon_e=\frac{3m_e^2}{2\omega_B} e^x}$, we see that this equation ultimately depends only on the logarithm of the energy
\begin{equation}
    \frac{\partial \tilde{\Phi}_e(x)}{\partial x}-\tilde{\Phi}_e(x)+\frac{1}{2}\tilde{\Phi}_e\left(\frac{x}{2}\right)=0.
\end{equation}
These equations are valid for $x<0$, since as we already discussed we only consider the regime where the energy of the radiated photon is lower than its parent lepton. For $|x|\gg 1$, the derivative term can be neglected, as confirmed by the solution we find, and $\tilde{\Phi}_e\sim -x^{-1}$; its derivative drops as $x^{-2}$ and is indeed negligible. Therefore, we finally find that the pair distribution in this energy range has the approximate form
\begin{equation}
    n_e(\varepsilon_e)=C'_e \frac{(\varepsilon_e/\varepsilon_{\gamma, \rm thr})^{-3}}{\ln\left(\frac{3m_e^2}{2\varepsilon_e \omega_B}\right)},
\end{equation}
with $C'_e$ a constant. 
The logarithmic dependence is weak, and therefore as an approximation we may simply take $n_e(\varepsilon_e)\simeq C_e (\varepsilon_e/\varepsilon_{\gamma,\rm thr})^{-3}$. This is the central result of this appendix, that we report in the main text, namely that in the soft-radiation regime the pair cascade approaches the spectrum $n_e(\varepsilon_e)\propto \varepsilon_e^{-3}$.

\section{Impact of non-monochromatic target photon fields on IC-dominated cascades}\label{sec:non_monochromatic_ic}

If the target photon field is non-monochromatic, the energy scale at which the absorption sets in will be determined by the highest-energy target photons $\varepsilon_{t,\rm max}$ -- so the threshold energy scale $\varepsilon_{\gamma, \rm thr}\simeq m_e^2/\varepsilon_{t, \rm max}$. However, if the target field decreases with energy sufficiently rapidly (the precise conditions will be elucidated below), the IC energy losses may be dominated by lower-energy photons, say around a scale $\varepsilon_{t,\rm min}$. In this case, the cascade comes from the photons splitting into pairs via $\gamma+\gamma_t\to e^++e^-$, but the pairs radiate most of their energy into photons with much lower energies, of the order of $\varepsilon_{t,\rm min}\gamma^2$. This is the same soft-radiation regime that we have identified in the synchrotron cascade: in fact, both the synchrotron energy losses and the typical frequency of the radiated synchrotron photons are in order of magnitude the same as for IC scattering from a photon field with frequency $\omega_B$. Therefore, the IC cascade in the soft-radiation regime can be obtained directly from Eq.~\ref{eq:synchrotron_cascade_em} by replacing $\omega_B\to \varepsilon_{t, \rm min}$ and $U_B\to U_t$, the energy density of the target photon field with energy $\varepsilon_{t, \rm min}$.

\begin{figure*}
    \includegraphics[width=\textwidth]{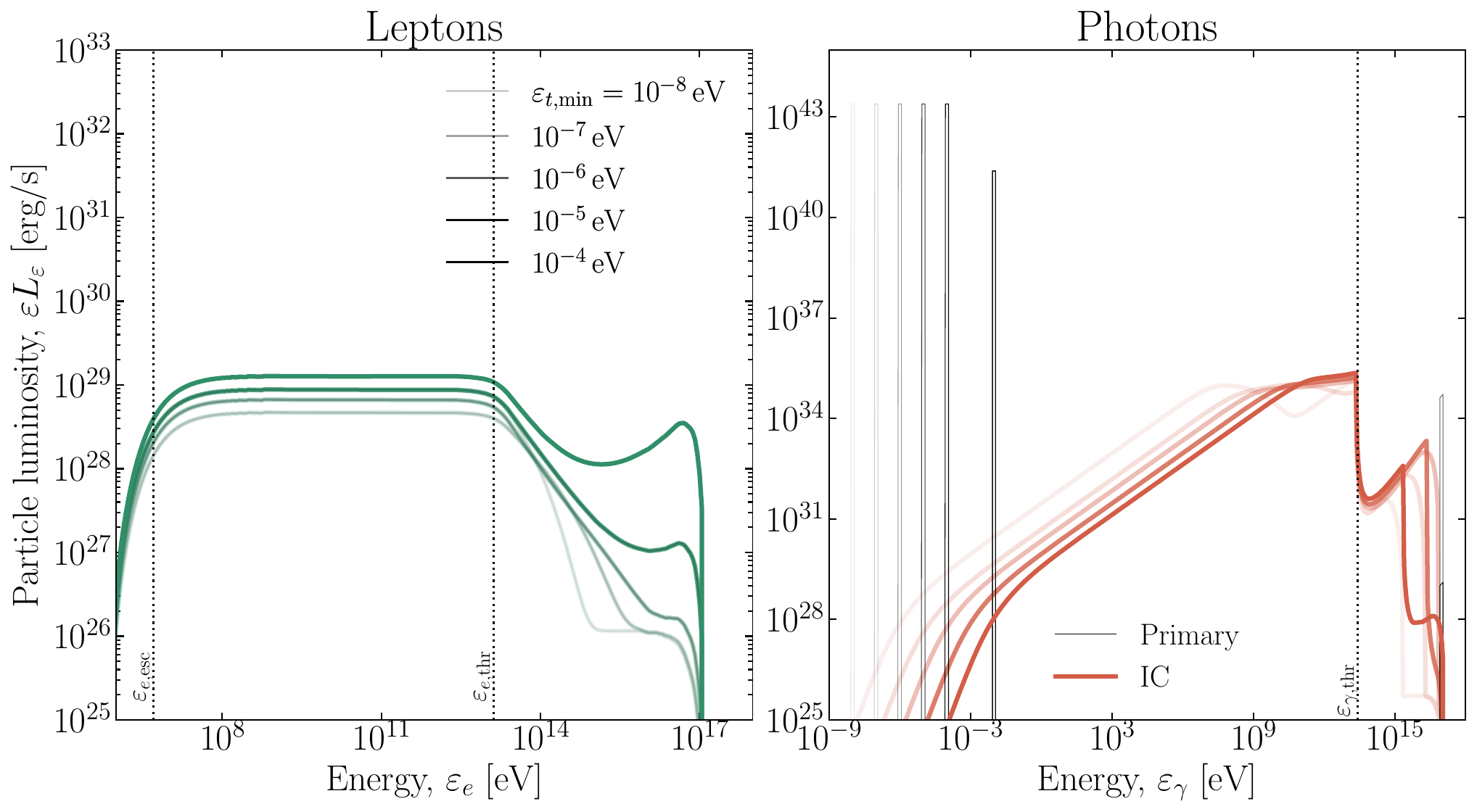}
\caption{\textbf{IC-dominated cascade for a bichromatic photon target.} Same as Fig.~\ref{fig:ic_cascade}, but with an additional photon field at a lower energy $\varepsilon_{t,\rm min}$. Different curve opacities refer to different values of $\varepsilon_{t,\rm min}$, which increases with increasing opacities. We do not highlight the regimes with different colors, since their energy ranges and nature depend on $\varepsilon_{t, \rm min}$. The pair cascade transitions from soft-radiation (at high energies) to pair-dominated (at low energies) regime at the threshold energy $\varepsilon_{e, \rm thr}$. This leads to to the characteristic broken-power-law photon spectrum; due to the varying target photon energy $\varepsilon_{t, \rm min}$, the position of the break changes with the curve opacity, approximately lying at $\varepsilon_{\gamma, \rm br}\simeq \varepsilon_{t, \rm min}(\varepsilon_{e, \rm thr}/m_e)^2$.}\label{fig:ic_bichromatic}
\end{figure*}

To make these statements more concrete, let us examine the simplest non-monochromatic target photon field, a bichromatic field. We consider the same setup as in Fig.~\ref{fig:ic_cascade}, with a monochromatic photon field at $\varepsilon_{t,\rm max}=10^{-2}\,\mathrm{eV}$ and a luminosity $L_t=10^{41}\,\mathrm{erg/s}$, but in addition we inject a second monochromatic field with $L_{t,\rm min}=10^{43}\,\mathrm{erg/s}$ at an energy $\varepsilon_{t,\rm min}$, which we leave as a varying parameter. The choice of $L_{t, \rm min}$ is driven by the requirement that the IC energy losses be dominated by the target photons at $\varepsilon_{t, \rm min}$; since in the Thomson regime the energy losses are proportional to $u_t$ (see Eq.~\ref{eq:ic_energy_loss_rate}), the energy density of the target photon field, this will naturally happen if $L_{t, \rm min}>L_{t, \rm max}$. The target photon spectrum is visible in Fig.~\ref{fig:ic_cascade} as the black line at low energies, which are injected as primaries; for increasing curve opacities, $\varepsilon_{t,\rm min}$ increases, so that the low-energy monochromatic field moves to higher energies.

The resulting cascade emission has all the qualitative features of the soft-radiation regime. In particular, the photon spectrum exhibits the typical break from $s_\gamma=2$ at high energies to $s_\gamma=3/2$ at low energies. The break happens, as anticipated, at an energy of about $\varepsilon_{\gamma,\rm br}\simeq \varepsilon_{t,\rm min} (\varepsilon_{e,\rm thr}/m_e)^2$. Indeed, we can clearly see that, as the opacity of the curve increases, the break moves to higher energies. 

Notice that the most transparent curve, corresponding to $\varepsilon_{t,\rm min}=10^{-8}\,\mathrm{eV}$, does not exhibit the typical $n_\gamma(\varepsilon_\gamma)\propto \varepsilon_\gamma^{-2}$. The reason is that $\varepsilon_{t,\rm min}$ is so low that the pairs injected at high energies, around $\varepsilon_{e, \rm he}=\varepsilon_{\gamma, \rm he}/2$, radiate at a much lower energy of the order of $\varepsilon_{t, \rm min} (\varepsilon_{\gamma, \rm he}/m_e)^2\ll \varepsilon_{e, \rm he}$. Thus, in the intermediate energy interval, there is no radiation injected; the pairs are actually in the cooling-only regime, rather than the soft-radiation regime, due to the absence of radiated photons, and therefore settle into a power law with $s_e=2$, clearly visible in the left panel of Fig.~\ref{fig:ic_bichromatic}. We have resolved to show an example with this somewhat subtle effect to provide a complete discussion; however, it appears that this effect is unlikely to show up in practical setups, and we do not find any such example in the astrophysical benchmark cases considered in Sec.~\ref{sec:astro_sources}, except marginally in our TDE benchmark. 

The appearance of the $n_\gamma(\varepsilon_\gamma)\propto \varepsilon_\gamma^{-2}$ spectrum in the presence of a non-monochromatic target photon field was in some sense the central feature of the Berezinsky theory of the electromagnetic cascade~\cite{Berezinsky:2016feh}. In that case, a bichromatic field was also used to derive it, which roughly simulates the effect of high-energy gamma-ray propagation in the target field of the extragalactic background light and the CMB, which lie at two very different energy scales. Our new insight is that the appearance of this spectrum is more generic, and is the signature of a soft-radiation regime which appears also in synchrotron dominated cascades; the latter are more phenomenologically relevant in the case of cascades developed inside high-energy astrophysical sources, as discussed in Sec.~\ref{sec:astro_sources}.

Under what conditions is a description in terms of two characteristic energies, similar to a bichromatic spectrum, appropriate? The propagation through an extragalactic photon field is of course a specially simple case, in which the target radiation is indeed composed of multiple thermal bumps for which a bichromatic approximation is relatively good. Within astrophysical sources, however, the target photon spectrum is often non-thermal and behaves as a power law over a wide energy range. In this case, the bichromatic representation is only appropriate if the IC energy losses are dominated by the low-energy part of the spectrum. Vice versa, if the IC energy losses are dominated by the high-energy part of the spectrum, we can simply replace the target photon spectrum with a monochromatic distribution around the highest energy of the target field. We will now prove that if the target field behaves as a power law $n_t(\varepsilon_t)\propto \varepsilon_t^{-s_t}$ in the energy range $\varepsilon_{t,\rm min}<\varepsilon_t<\varepsilon_{t,\rm max}$, then for $s_t<0$ one can replace it with a monochromatic target field at $\varepsilon_{t, \rm max}$, while for $s_t>2$ one can instead replace it with a bichromatic target field, with the low-energy part around $\varepsilon_{t,\rm min}$ dominating the IC losses, while the high-energy part around $\varepsilon_{t, \rm max}$ makes the environment optically thick to $\gamma\gamma$ interactions. The intermediate cases depend on the specific spectral index of the target field, and therefore lose the universality that we have otherwise identified. 

If the power law is sufficiently hard, then most of the energy is concentrated around $\varepsilon_{t,\rm max}$. In this case, one can expect the monochromatic approximation to be relatively accurate, and replace the power-law spectrum with a monochromatic target at $\varepsilon_{t, \rm max}$. In order for this to be a good approximation, the IC interaction rate should be dominated by the interaction with the highest energy target for any energy of the lepton. In this case, the interaction rate of a lepton with energy $\varepsilon_e$ is, in order of magnitude,
\begin{equation}
    \Gamma_{\rm IC}(\varepsilon_e)\simeq \int_{m_e^2/\varepsilon_e}^{\varepsilon_{t, \rm max}}d\varepsilon_t n_t(\varepsilon_t) \sigma_T \frac{m_e^2}{\varepsilon_t \varepsilon_e};
\end{equation}
the last factor is the Klein-Nishina suppression, which appears for $\varepsilon_t \varepsilon_e\gtrsim m_e^2$ (justifying the choice of the lower bound of integration). In order for this integral to be dominated by its upper bound, we must have $s_t<0$. Thus, for $s_t<0$, the monochromatic approximation is an appropriate one.

If the power law is instead very soft, we can have the opposite situation in which the IC interaction rate is dominated by the low-energy target photons, around $\varepsilon_{t,\rm min}$, for any lepton energy. In this case, the bichromatic approximation is appropriate, with the target photons at $\varepsilon_{t,\rm min}$ dominating the IC losses, while the target photons at $\varepsilon_{t, \rm max}$ cause $\gamma\gamma$ absorption. Therefore, we should now determine when are IC losses dominated by the low-energy target photons. The interaction with these low-energy photons is in the Thomson regime, in which the electron-photon cross section is approximately of the order of $\sigma_T$; however, the fraction of energy lost by the electron in each scattering is of the order of $\varepsilon_e \varepsilon_t/m_e^2$. Therefore, in this case the interaction rate reads
\begin{equation}
    \Gamma_{\rm IC}(\varepsilon_e)\simeq \int_{\varepsilon_{t,\rm min}}^{m_e^2/\varepsilon_e}d\varepsilon_t n_t(\varepsilon_t)\sigma_T \frac{\varepsilon_t \varepsilon_e}{m_e^2}.
\end{equation}
In order for this integral to be dominated by its lower bound, we must have $s_t>2$. Under these conditions, the target photon spectrum can qualitatively be replaced by a bichromatic approximation.
In the intermediate cases, with $0<s_t<2$, the IC energy loss rate depends on the specific properties of the target photon spectrum. Therefore, a universal shape for the cascade cannot be recovered; its properties will unavoidably depend on the specific setup considered.

\section{Inverse-Compton dominated cascades in GRBs}\label{app:ic_dominated_grb}

\begin{figure*}
    \includegraphics[width=\textwidth]{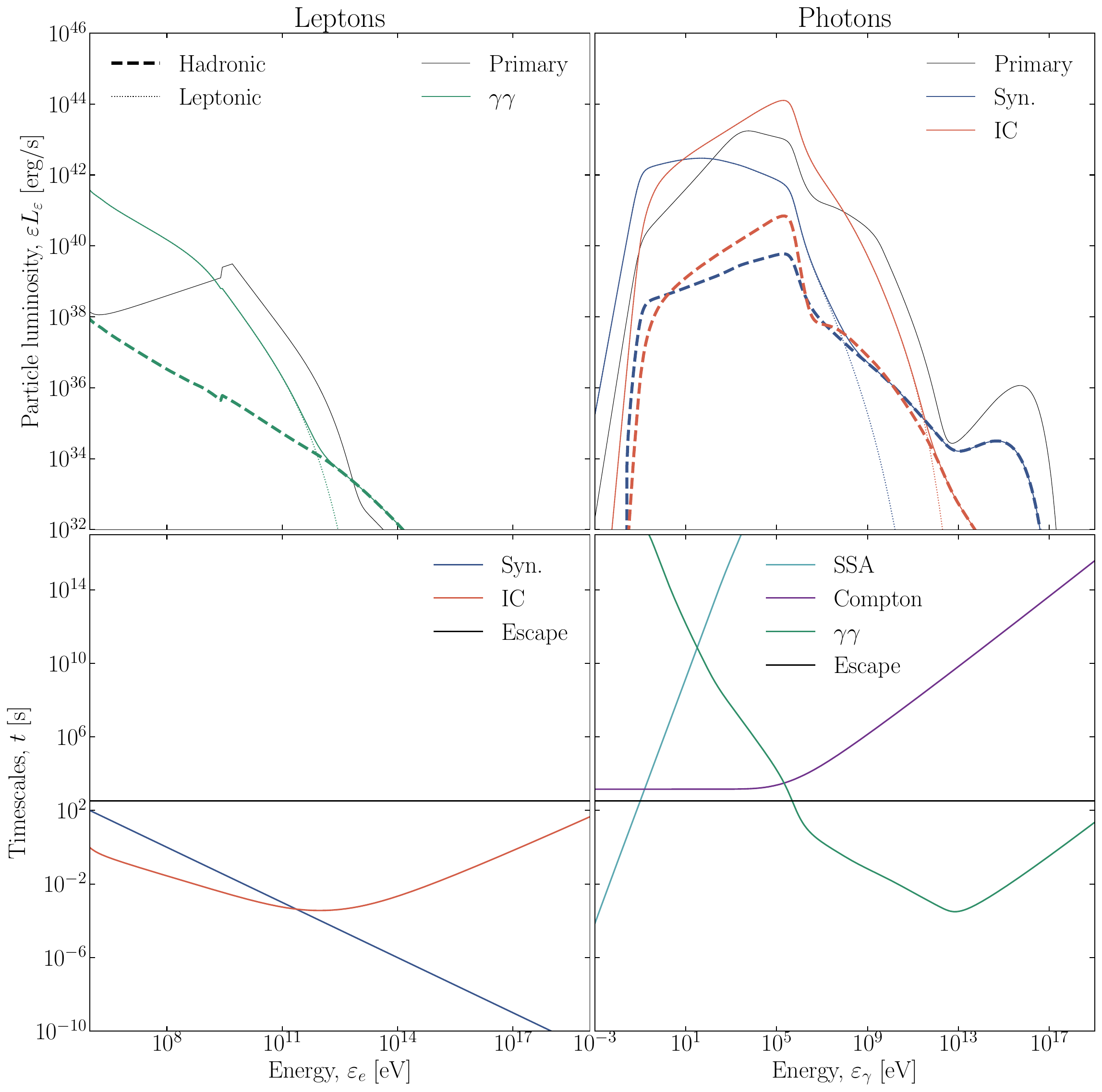}
    \caption{\textbf{Electromagnetic cascade in a GRB-inspired lepto-hadronic model with IC-dominant energy losses.} Same as Fig.~\ref{fig:grb}, for an IC-dominated case with $L_e=10^{45}\,\mathrm{erg/s}$. The numerical values for the other parameters are the same as in Table~\ref{tab:grb_parameters}. The resulting cascade spectrum is not universal, since the IC energy losses depend on the lepton energy in a way that is sensitive to the target photon spectrum.}\label{fig:grb_com}
\end{figure*}

For completeness, we can also consider a GRB-inspired case in which the cascade is IC-dominated. Since the GRB target photon spectrum is entirely non-thermal, based on our discussion in Appendix~\ref{sec:non_monochromatic_ic}, we do not expect the resulting cascade to exhibit the universal behavior we have identified. We are now going to confirm this statement by an explicit numerical example. We adopt the same benchmark parameters as in Table~\ref{tab:grb_parameters}, but increase the lepton luminosity to $L_e=10^{45}\,\mathrm{erg/s}$, so that the IC losses off the radiation from the leptons is significantly enhanced.

The resulting emission is shown in Fig.~\ref{fig:grb_com}. Due to the much larger lepton luminosity, the signal is now entirely dominated by the leptonic cascade, peaking at about $\varepsilon_\gamma\sim 10^5\,\mathrm{eV}$. When the hadronic component is injected (shown in dashed), the pairs produced by $\gamma\gamma$ absorption settle into a power-law shape; however, this power law is not of the form of our universal cascade prediction, in agreement with our expectation. In fact, this breaking of universality is even more plainly visible from the energy dependence of the timescale for IC losses, in the bottom left panel; below $\varepsilon_e=10^{11}\,\mathrm{eV}$, the timescale does not decrease as $t_{\rm IC}\propto \varepsilon_e^{-1}$ as one would expect in the Thomson regime. Instead, its energy dependence depends on the target photon spectrum, which immediately hinders the emergence of a universal cascade spectrum.

\bibliographystyle{bibi}
\bibliography{References}

\end{document}